\documentclass[english,aps,prd,twocolumn,amsmath,superscriptaddress,showpacs]{revtex4-1}
\usepackage[T1]{fontenc}
\usepackage[latin1]{inputenc}
\usepackage{longtable}
\usepackage{graphicx}

\usepackage{url}
\usepackage{xspace}
\usepackage{multirow}
\usepackage{units}
\usepackage[normalem]{ulem}
\usepackage{color}

\newcommand{\ppbar}{\ensuremath{p\bar{p}}\xspace}
\newcommand{\ttbar}{\ensuremath{t\bar{t}}\xspace}
\newcommand{\ljets}{\ensuremath{\ell}+{\rm jets}\xspace}
\newcommand{\wjets}{\ensuremath{W}+{\rm jets}\xspace}
\newcommand{\pt}{\ensuremath{p_{T}}\xspace}
\newcommand{\ptmiss}{\ensuremath{p \kern-0.5em\slash_{T}}\xspace}
\newcommand{\mtop}{\ensuremath{m_{t}}\xspace}
\newcommand{\mtgen}{\ensuremath{m_{t}^{\rm gen}}\xspace}
\newcommand{\kjes}{\ensuremath{k_{\rm JES}}\xspace}
\newcommand{\kjgen}{\ensuremath{k_{\rm JES}^{\rm gen}}\xspace}
\newcommand{\xt}{\ensuremath{\tilde x}\xspace}
\newcommand{\alpgen}{{\sc alpgen}\xspace}
\newcommand{\pythia}{{\sc pythia}\xspace}
\newcommand{\vecbos}{{\sc vecbos}\xspace}
\newcommand{\mcnlo}{{\sc mc@nlo}\xspace}
\newcommand{\herwig}{{\sc herwig}\xspace}
\newcommand{\mcfm}{{\sc mcfm}\xspace}
\newcommand{\geant}{{\sc geant3}\xspace}
\newcommand{\sigmat}{\ensuremath{\tilde\sigma}\xspace}

\newcommand{\Psig}{\ensuremath{P_{\rm sig}}\xspace}
\newcommand{\Pbkg}{\ensuremath{P_{\rm bkg}}\xspace}
\newcommand{\etal}{\emph{et al.}\xspace}
\usepackage{babel}
\begin{document}

\hspace{5.2in} \mbox{FERMILAB-PUB-11-240-E}
\title{Precise measurement of the top-quark mass from lepton$+$jets events
at D0}

\affiliation{Universidad de Buenos Aires, Buenos Aires, Argentina}
\affiliation{LAFEX, Centro Brasileiro de Pesquisas F{\'\i}sicas, Rio de Janeiro, Brazil}
\affiliation{Universidade do Estado do Rio de Janeiro, Rio de Janeiro, Brazil}
\affiliation{Universidade Federal do ABC, Santo Andr\'e, Brazil}
\affiliation{Instituto de F\'{\i}sica Te\'orica, Universidade Estadual Paulista, S\~ao Paulo, Brazil}
\affiliation{Simon Fraser University, Vancouver, British Columbia, and York University, Toronto, Ontario, Canada}
\affiliation{University of Science and Technology of China, Hefei, People's Republic of China}
\affiliation{Universidad de los Andes, Bogot\'{a}, Colombia}
\affiliation{Charles University, Faculty of Mathematics and Physics, Center for Particle Physics, Prague, Czech Republic}
\affiliation{Czech Technical University in Prague, Prague, Czech Republic}
\affiliation{Center for Particle Physics, Institute of Physics, Academy of Sciences of the Czech Republic, Prague, Czech Republic}
\affiliation{Universidad San Francisco de Quito, Quito, Ecuador}
\affiliation{LPC, Universit\'e Blaise Pascal, CNRS/IN2P3, Clermont, France}
\affiliation{LPSC, Universit\'e Joseph Fourier Grenoble 1, CNRS/IN2P3, Institut National Polytechnique de Grenoble, Grenoble, France}
\affiliation{CPPM, Aix-Marseille Universit\'e, CNRS/IN2P3, Marseille, France}
\affiliation{LAL, Universit\'e Paris-Sud, CNRS/IN2P3, Orsay, France}
\affiliation{LPNHE, Universit\'es Paris VI and VII, CNRS/IN2P3, Paris, France}
\affiliation{CEA, Irfu, SPP, Saclay, France}
\affiliation{IPHC, Universit\'e de Strasbourg, CNRS/IN2P3, Strasbourg, France}
\affiliation{IPNL, Universit\'e Lyon 1, CNRS/IN2P3, Villeurbanne, France and Universit\'e de Lyon, Lyon, France}
\affiliation{III. Physikalisches Institut A, RWTH Aachen University, Aachen, Germany}
\affiliation{Physikalisches Institut, Universit{\"a}t Freiburg, Freiburg, Germany}
\affiliation{II. Physikalisches Institut, Georg-August-Universit{\"a}t G\"ottingen, G\"ottingen, Germany}
\affiliation{Institut f{\"u}r Physik, Universit{\"a}t Mainz, Mainz, Germany}
\affiliation{Ludwig-Maximilians-Universit{\"a}t M{\"u}nchen, M{\"u}nchen, Germany}
\affiliation{Fachbereich Physik, Bergische Universit{\"a}t Wuppertal, Wuppertal, Germany}
\affiliation{Panjab University, Chandigarh, India}
\affiliation{Delhi University, Delhi, India}
\affiliation{Tata Institute of Fundamental Research, Mumbai, India}
\affiliation{University College Dublin, Dublin, Ireland}
\affiliation{Korea Detector Laboratory, Korea University, Seoul, Korea}
\affiliation{CINVESTAV, Mexico City, Mexico}
\affiliation{FOM-Institute NIKHEF and University of Amsterdam/NIKHEF, Amsterdam, The Netherlands}
\affiliation{Radboud University Nijmegen/NIKHEF, Nijmegen, The Netherlands}
\affiliation{Joint Institute for Nuclear Research, Dubna, Russia}
\affiliation{Institute for Theoretical and Experimental Physics, Moscow, Russia}
\affiliation{Moscow State University, Moscow, Russia}
\affiliation{Institute for High Energy Physics, Protvino, Russia}
\affiliation{Petersburg Nuclear Physics Institute, St. Petersburg, Russia}
\affiliation{Instituci\'{o} Catalana de Recerca i Estudis Avan\c{c}ats (ICREA) and Institut de F\'{i}sica d'Altes Energies (IFAE), Barcelona, Spain}
\affiliation{Stockholm University, Stockholm and Uppsala University, Uppsala, Sweden}
\affiliation{Lancaster University, Lancaster LA1 4YB, United Kingdom}
\affiliation{Imperial College London, London SW7 2AZ, United Kingdom}
\affiliation{The University of Manchester, Manchester M13 9PL, United Kingdom}
\affiliation{University of Arizona, Tucson, Arizona 85721, USA}
\affiliation{University of California Riverside, Riverside, California 92521, USA}
\affiliation{Florida State University, Tallahassee, Florida 32306, USA}
\affiliation{Fermi National Accelerator Laboratory, Batavia, Illinois 60510, USA}
\affiliation{University of Illinois at Chicago, Chicago, Illinois 60607, USA}
\affiliation{Northern Illinois University, DeKalb, Illinois 60115, USA}
\affiliation{Northwestern University, Evanston, Illinois 60208, USA}
\affiliation{Indiana University, Bloomington, Indiana 47405, USA}
\affiliation{Purdue University Calumet, Hammond, Indiana 46323, USA}
\affiliation{University of Notre Dame, Notre Dame, Indiana 46556, USA}
\affiliation{Iowa State University, Ames, Iowa 50011, USA}
\affiliation{University of Kansas, Lawrence, Kansas 66045, USA}
\affiliation{Kansas State University, Manhattan, Kansas 66506, USA}
\affiliation{Louisiana Tech University, Ruston, Louisiana 71272, USA}
\affiliation{Boston University, Boston, Massachusetts 02215, USA}
\affiliation{Northeastern University, Boston, Massachusetts 02115, USA}
\affiliation{University of Michigan, Ann Arbor, Michigan 48109, USA}
\affiliation{Michigan State University, East Lansing, Michigan 48824, USA}
\affiliation{University of Mississippi, University, Mississippi 38677, USA}
\affiliation{University of Nebraska, Lincoln, Nebraska 68588, USA}
\affiliation{Rutgers University, Piscataway, New Jersey 08855, USA}
\affiliation{Princeton University, Princeton, New Jersey 08544, USA}
\affiliation{State University of New York, Buffalo, New York 14260, USA}
\affiliation{Columbia University, New York, New York 10027, USA}
\affiliation{University of Rochester, Rochester, New York 14627, USA}
\affiliation{State University of New York, Stony Brook, New York 11794, USA}
\affiliation{Brookhaven National Laboratory, Upton, New York 11973, USA}
\affiliation{Langston University, Langston, Oklahoma 73050, USA}
\affiliation{University of Oklahoma, Norman, Oklahoma 73019, USA}
\affiliation{Oklahoma State University, Stillwater, Oklahoma 74078, USA}
\affiliation{Brown University, Providence, Rhode Island 02912, USA}
\affiliation{University of Texas, Arlington, Texas 76019, USA}
\affiliation{Southern Methodist University, Dallas, Texas 75275, USA}
\affiliation{Rice University, Houston, Texas 77005, USA}
\affiliation{University of Virginia, Charlottesville, Virginia 22901, USA}
\affiliation{University of Washington, Seattle, Washington 98195, USA}
\author{V.M.~Abazov} \affiliation{Joint Institute for Nuclear Research, Dubna, Russia}
\author{B.~Abbott} \affiliation{University of Oklahoma, Norman, Oklahoma 73019, USA}
\author{B.S.~Acharya} \affiliation{Tata Institute of Fundamental Research, Mumbai, India}
\author{M.~Adams} \affiliation{University of Illinois at Chicago, Chicago, Illinois 60607, USA}
\author{T.~Adams} \affiliation{Florida State University, Tallahassee, Florida 32306, USA}
\author{G.D.~Alexeev} \affiliation{Joint Institute for Nuclear Research, Dubna, Russia}
\author{G.~Alkhazov} \affiliation{Petersburg Nuclear Physics Institute, St. Petersburg, Russia}
\author{A.~Alton$^{a}$} \affiliation{University of Michigan, Ann Arbor, Michigan 48109, USA}
\author{G.~Alverson} \affiliation{Northeastern University, Boston, Massachusetts 02115, USA}
\author{G.A.~Alves} \affiliation{LAFEX, Centro Brasileiro de Pesquisas F{\'\i}sicas, Rio de Janeiro, Brazil}
% ...
%
%
\collaboration{The D0 Collaboration\footnote{with visitors from
$^{a}$Augustana College, Sioux Falls, SD, USA,
$^{b}$The University of Liverpool, Liverpool, UK,
$^{c}$SLAC, Menlo Park, CA, USA,
$^{d}$University College London, London, UK,
$^{e}$Centro de Investigacion en Computacion - IPN, Mexico City, Mexico,
$^{f}$ECFM, Universidad Autonoma de Sinaloa, Culiac\'an, Mexico,
and 
$^{g}$Universit{\"a}t Bern, Bern, Switzerland.
}} \noaffiliation
\vskip 0.25cm

\date{July 5, 2011}

\begin{abstract}
We report a measurement of the mass of the top quark in lepton$+$jets final
states of $p\overline{p}\rightarrow$\ttbar data corresponding to $2.6$ fb$^{-1}$
of integrated luminosity collected by the D0 experiment at the Fermilab
Tevatron Collider.  
A matrix-element method is developed that combines an {\sl in situ} jet energy
calibration with our standard jet energy scale derived from studies of
$\gamma+$jet and dijet events. We then implement a flavor-dependent
jet response correction through a novel approach.
This method is used to measure a top-quark mass of $\mtop=176.01\pm1.64$ GeV.
Combining this result with our previous result obtained on an independent data set, 
we measure a top-quark mass of $\mtop=174.94\pm 1.49$ GeV for a total integrated
luminosity of $3.6$ fb$^{-1}$.
\end{abstract}
\maketitle
\section{Introduction}
The observation of the top quark in
1995~\citep{topdiscovery_dzero,topdiscovery_cdf} confirmed the  existence of the
six quarks in three generations of fermions expected in the  standard model (SM)
of particle interactions. 
Because of its mass, the lifetime of the top quark is much shorter than 
the time-scale of hadronization.
The top quark can therefore decay before interacting, making it the only 
quark whose characteristics can be studied in isolation.
The large mass of the top quark (\mtop), corresponding to a Yukawa coupling to the Higgs
boson equal to $1$ within the current uncertainties, suggests a special role for
the top quark in the breaking of electroweak symmetry. It is therefore not
surprising that the precise determination of the mass of the top quark has
received great attention.  The interest in the top-quark mass also arises from
the constraint  imposed on the mass of the Higgs boson, $m_H$, from the
relationship among the values  of \mtop, $m_H$, and the SM radiative corrections
to the mass of the $W$ boson~\citep{lepewwg}. A precise measurement of \mtop
also provides a useful constraint on contributions from physics beyond the
standard model~\citep{bsm}.

The statistical uncertainty on the world average value of \mtop is 0.3\%, and
the accuracy of the measurement of \mtop is now dominated by systematic 
uncertainties~\citep{tevewwg}. The main systematic contributions arise from
uncertainties on the jet energy calibration and on the Monte Carlo (MC)
simulation of \ttbar  events.

We present a new measurement of the mass of the top quark based on
$2.6$~fb$^{-1}$  of integrated luminosity from $p\overline{p}$ collisions at
$\sqrt{s}=1.96$~TeV, collected with the D0 detector at the Fermilab Tevatron
Collider.  The analysis  focuses on \ttbar events identified in lepton$+$jets
(\ljets) final  states (with $\ell$ representing either an electron or a
muon)~\citep{taulepton}, in which the top  and antitop quark are assumed to
decay into a $W$ boson and $b$ quark \citep{smtopdecay}, with one of the $W$
bosons in the $W^{+}W^{-}b\overline{b}$ final system decaying  via
$W\rightarrow\ell\nu$ into a lepton and neutrino and the other via $W\rightarrow
q\overline{q}^{\prime}$  into two quarks, and all four quarks
($q\overline{q}^{\prime}b\overline{b}$) hadronizing into jets.  Such events are
characterized by an isolated electron or muon with large transverse momentum
($p_{T}$), an undetected neutrino that causes a  large imbalance in transverse
momentum, and four high-$p_{T}$ jets.  In selecting  candidate events, we
exploit this distinct signature, which helps distinguish  these events from
background.  

Compared to the previous measurement based on data corresponding to
$1$~fb$^{-1}$ of integrated luminosity~\citep{meljprl}, we use a larger data set
and an improved evaluation of systematic  uncertainties.  The analysis uses the
same matrix element (ME) analysis technique, with an {\sl in situ} jet energy
calibration based on constraining the invariant mass  of the two jets from the
decay of the $W$ boson to the world average  value of  $M_W=80.4$
GeV~\citep{pdg}. As in the previous measurement, the standard jet energy scale
(JES),  derived from $\gamma+$jet and dijet data samples, is used as an
additional  constraint, and implemented through a Gaussian prior on its absolute
value and uncertainty. A major improvement in this new measurement is the
significant reduction of the  uncertainty associated with the modeling of
differences in the calorimeter response to $b$-quark and light-quark jets
originating from the introduction of a new flavor-dependent jet energy response
correction.

This measurement, like all direct measurements of \mtop, relies on MC \ttbar
events for absolute calibration.  It is therefore important to understand the
precise definition of the input mass \mtgen in MC \ttbar event generators, such
as \alpgen~\citep{alpgen} and \pythia~\citep{pythia}, used to calibrate the
direct measurements.   Although \mtgen is not well defined in leading order (LO)
generators that use parton showers to model higher-order effects and
hadronization, it has been argued that \mtgen should be viewed as being close to
the pole mass~\citep{buckley}.  In Ref. \citep{xsmass}, the D0 Collaboration has
extracted \mtop from a comparison of the measured \ttbar production cross
section with predictions from higher-order quantum chromodynamics (QCD), by
equating \mtgen both with the pole mass ($\mtop^{\rm pole}$) and with the
$\overline{\rm MS}$ mass ($\mtop^{\rm \overline{MS}}$).  The extracted \mtop,
under the assumption $\mtgen\equiv\mtop^{\rm pole}$, is found to agree with the
average value of \mtop from the Tevatron, while the \mtop extracted assuming
$\mtgen\equiv\mtop^{\rm \overline{MS}}$ is found to be different from the
average value of \mtop.  These results favor the pole mass interpretation of
\mtgen.

This paper is arranged as follows.  A brief description of the D0 detector is
given in Sec. \ref{sec:detector},  which is followed by a discussion of the
selection and reconstruction of the physical objects in this analysis in Sec.
\ref{sec:eventsel}. Section \ref{sec:mcsamples} summarizes the MC samples used
to simulate the events of interest, and Sec.~\ref{sec:me_method} discusses the
technique used to extract the value of \mtop. This is followed by a description
of the calibration of the response of the analysis method in
Sec.~\ref{sec:calib} and a discussion of the flavor-dependent jet response
correction used to bring the simulation of the calorimeter response to jets into
agreement with data in Sec.~\ref{sec:flavcorr}. The result of the calibration is
applied to the data in Sec. \ref{sec:results}, where the measured value of \mtop
and its statistical uncertainty are also presented. Section \ref{sec:syst} 
describes the evaluation of systematic uncertainties and the final result is
given in Sec.~\ref{sec:currentresult}.  We combine this new measurement in
Sec.~\ref{sec:combination} with an updated version of that from Ref.
\citep{meljprl} in which the flavor-dependent jet response correction mentioned
above has been applied and the systematic uncertainties have been updated.

\section{The D0 detector}
\label{sec:detector}
The D0 detector consists primarily of a magnetic central tracking system,
calorimetry, and a muon system.  The central tracking system comprises a
silicon microstrip tracker (SMT) and a central fiber  tracker (CFT), both
located within a 1.9~T superconducting solenoidal  magnet~\cite{run2det}. The
SMT~\cite{run2smt} has $\approx 800,000$ individual strips,  with typical pitch
of $50-80$ $\mu$m, and a design optimized for track and vertex finding at
$|\eta|<2.5$, where the pseudorapidity $\eta=-\ln\left[\tan(\theta/2)\right]$,
and $\theta$ is the polar angle with respect to the proton beam direction
relative to the center of the detector.  The system has a six-barrel
longitudinal structure, each with a set  of four layers arranged axially around
the beam pipe, and interspersed  with 16 radial disks.  In 2006, a fifth layer,
referred to as \emph{Layer 0}, was installed close to the beam
pipe~\citep{run2ab,run2lyr0}.  The CFT has eight thin coaxial barrels, each 
supporting two doublets of overlapping scintillating fibers of 0.835~mm 
diameter, one doublet being parallel to the collision axis, and the  other
alternating by $\pm 3^{\circ}$ relative to the axis. Light signals  are
transferred via clear fibers to solid-state photon counters (VLPCs)  that have
$\approx 80$\% quantum efficiency.

Central and forward preshower detectors, located just outside of the
superconducting coil (in front of the calorimetry), are constructed of several
layers of extruded triangular scintillator strips that are read  out using
wavelength-shifting fibers and the VLPC.  These detectors provide initial
sampling of electromagnetic showers, and thereby help distinguish incident
photons from electrons.  The next layer of  detection involves three
liquid-argon/uranium calorimeters: a central  section (CC) covering $|\eta|$ up
to $\approx 1.1$, and two end calorimeters (EC) that extend coverage to
$|\eta|\approx 4.2$, all  housed in separate cryostats.  The electromagnetic
(EM) section of the calorimeter is segmented into four layers, with transverse
segmentation of the cells in pseuodorapidity and azimuth of
$\Delta\eta\times\Delta\phi=0.1\times0.1$, except for the third layer, where the
segmentation is $0.05\times0.05$.  The hadronic portion of the calorimeter is
located after the EM sections, and consists of fine hadron-sampling layers,
followed by more coarse hadronic layers.  In addition, scintillators between the
CC and EC cryostats provide  sampling of developing showers for
$1.1<|\eta|<1.4$.

A muon system~\cite{run2muon} is located beyond the calorimetry, and consists of
a  layer of tracking detectors and scintillation trigger counters  before 1.9~T
toroids, followed by two similar layers after the toroids. Tracking for
$|\eta|<1$ relies on 10~cm wide drift tubes, while 1~cm mini-drift tubes are
used for $1<|\eta|<2$.

Luminosity is measured using plastic scintillator arrays located in front  of
the EC cryostats, covering $2.7 < |\eta| < 4.4$.  The trigger and data
acquisition systems are designed to accommodate  the high instantaneous
luminosities of the Tevatron~\citep{run2det,run2l1cal}. Based on preliminary
information from  tracking, calorimetry, and muon systems, the output of the
first level  of the trigger is used to limit the rate for accepted events to 
$\approx$ 2~kHz.  At the next trigger stage, with more refined  information, the
rate is reduced further to $\approx$ 1~kHz. These first two levels of triggering
rely mainly on hardware and firmware. The third and final level of the trigger,
with access to all of the event  information, uses software algorithms and a
computing farm, and reduces  the output rate to $\approx$ 100~Hz, which is
written to tape.

\section{Object Reconstruction and Event Selection}
\label{sec:eventsel}

In the following sections, we summarize how the physical objects in data and MC
events are reconstructed from information in the detector and the criteria
applied to these objects to select the \ljets \ttbar candidate events.

\subsection{Object Reconstruction}

This section describes the reconstruction of electrons, muons, missing
transverse momentum, and jets, and the identification of $b$ jets.

\subsubsection{Identification of Electrons}

Electron candidates are defined by narrow clusters of energy deposited in towers
of the electromagnetic calorimeter located within a cone of radius ${\cal
R}=\sqrt{(\Delta \eta)^2+(\Delta \phi)^2}=0.2$, where $\Delta \eta$ is the
pseudorapidity, and $\Delta \phi$ is the azimuthal angle of each cluster
relative to the seed cluster.  At least 90\% of the total energy measured within
this cone is required to be located within the electromagnetic section to be
consistent with expectations for electromagnetic showers.  Isolation from energy
deposited by hadrons is imposed by requiring  $(E_{\rm tot}-E_{\rm EM })/E_{\rm
EM }<0.15$, where $E_{\rm tot }$ ($E_{\rm EM }$) is the total (electromagnetic)
energy in a cone of radius ${\cal R}=0.4$ (${\cal R}=0.2$).  Candidate electrons
are required to have longitudinal and transverse shower profiles compatible with
those of electromagnetic showers and to be spatially matched to a track
reconstructed in the central tracking system.  Electron candidates meeting these
criteria are referred to as \emph{loose} electrons.  Finally, (i) the value of a
multivariable likelihood discriminant based on tracking system and calorimeter
information is required to be consistent with that for an electron, and (ii) a
neural network, trained using information from the tracking system, calorimeter,
and central preshower detector is used to further reject background from jets
misidentified as electrons.  Electron candidates meeting these criteria are
referred to as \emph{tight} electrons, and are those used to obtain the final
selection.

\subsubsection{Identification of Muons}

Muons are identified by requiring a minimum number of wire and scintillator hits
on both sides of the toroidal magnets in the muon detector~\citep{run2muon}. 
Cosmic ray background is rejected by requiring scintillator signals consistent
in time with muons originating from the \ppbar collision.  Tracks in the muon
system are required to match a reconstructed track in the central tracker having
a small impact parameter with respect to the \ppbar interaction vertex (PV) to
reject muons from cosmic rays and decays in flight of kaons and pions.  Muon
candidates must also be isolated from jets with $\pt>15$ GeV by requring a
separation in $\eta-\phi$ space between the muon and jet of $\Delta {\cal
R}(\mu,{\rm jet})>0.5$~\citep{deltaR}.  Candidates satisfying these requirements
are referred to as \emph{loosely isolated} muons.  The following two variables
are used to impose additional isolation requirements:  $E_{\rm halo}^{\rm
scaled}$ is defined as the ratio of calorimeter energy within an annulus of
$0.1<{\cal R}<0.4$ around the muon direction to the $p_T$ of the muon; 
$p_{T,{\rm cone}}^{\rm scaled}$ is defined as the ratio of the total $p_T$ of
all tracks within a cone of ${\cal R}=0.5$, excluding the muon, to the $p_T$ of
the muon.  Muon candidates meeting all the requirements above that satisfy
$E_{\rm halo}^{\rm scaled}<0.12$ and $p_{T,{\rm cone}}^{\rm scaled}<0.12$ are
referred to as \emph{veto} muons.  Further tightening these requirements to
$E_{\rm halo}^{\rm scaled}<0.08$ and $p_{T,{\rm cone}}^{\rm scaled}<0.06$
selects candidates referred to as \emph{tightly isolated} muons.

\subsubsection{Measurement of the Imbalance in Transverse Momentum}

We use the conservation of momentum to measure the momentum imbalance in the
transverse plane (\ptmiss).  From that, we infer the presence of the neutrino. 
The \ptmiss is determined from the vector sum of the energies of all
cells in the electromagnetic and hadronic calorimeters. Subsequent energy
corrections applied to reconstructed objects such as jets and muons are also
propagated to the missing transverse momentum.

\subsubsection{Identification of Jets}

Jet candidates are reconstructed using the iterative midpoint cone algorithm
with a cone radius of ${\cal R}=0.5$~\citep{run2jets}.  Only calorimeter cells
with energies that are $2.5$ standard deviations above the mean of the noise
distribution are considered in the reconstruction.  Isolated cells with energies
less than $4$ standard deviations above the mean of the noise distribution are
also discarded.  Among the jet candidates with $\pt>8$ GeV, the following
selection criteria are imposed. The electromagnetic fraction of the jet energy
is required to be below $0.95$ to reject electrons and above $0.05$ to suppress
jets dominated by noise from the hadronic part of the calorimeter.  Jets with a
large fraction of their energy deposited in the coarse hadronic layers of the
calorimeter are rejected to suppress jets dominated by noise typical for those
layers.  To minimize background from jet candidates arising from noise in the
precision readout of the calorimeter, confirmation from the readout system of
the first level trigger is required for reconstructed jets.  Jets matched to
loose electrons with $\pt>20$ GeV and $\Delta {\cal
R}(e,\rm{jet})<0.5$~\citep{deltaR} are also rejected.  Energies of jets
containing muons are corrected with the measured muon momentum after accounting
for the typical energy deposited by a minimum ionizing particle.

The energy of a reconstructed jet is corrected, on average, to that of a
\emph{particle jet}~\citep{partjet} containing the final-state particles within
a cone of radius ${\cal R}=0.5$  corresponding to the reconstructed jet.  The
first step involves the subtraction of the offset energy due to calorimeter
noise and contributions from previous and following beam crossings and multiple
interactions within the same beam crossing.  This is followed by an absolute
response correction determined from $\gamma+$jet events and a relative
$\eta$-dependent correction based on $\gamma+$jet and dijet events.  Finally, a
showering correction is applied to account for the lateral leakage of energy
across the jet cone boundary.

\subsubsection{Identification of $b$ jets}
The lifetime of the $b$ quark, unlike that of the top quark, is far longer than 
the time-scale for hadronization. This means that, during QCD  evolution,
the $b$ quark can form short lived $b$ hadrons that travel $\gtrsim 1$~mm before
decaying through the weak interaction.  We identify the $b$ jets among the
candidates satisfying the jet selection criteria described in the previous
section by using a neural network (NN) $b$-tagging algorithm that selects jets
with displaced vertices and tracks relative to the PV~\citep{bnim}.  The NN
tagger is based on nine input variables that can be separated into two
categories.  The first category is related to the reconstructed secondary vertex
and includes the vertex quality, the number of associated tracks, the invariant
mass of the vertex, the number of secondary vertices reconstructed within the
jet, the spatial separation between the jet axis and the position vector of the
secondary vertex relative to the PV, and the length of the flight path projected
on the transverse plane divided by its uncertainty (which provides a measure of
the decay length significance in terms of standard deviations).  The second
category relies only on the characteristics of the tracks within the jets such
as impact parameters, transverse momentum, and track quality. The $b$-jet
candidates are also required to have at least two good quality tracks
originating from the PV. The tagging efficiency for $b$ jets is $\approx 65\%$
for a misidentification rate of $\approx 3\%$ for $u$, $d$, $s$ quark, or gluon
jets~\citep{btageff}.

\subsection{Event Selection}
The data sample used in this analysis was collected with the D0 detector at the
Tevatron between June 2006 and June 2008, and corresponds to an integrated
luminosity of 2.6 fb$^{-1}$.  The selected events must satisfy a single-lepton
trigger, requiring a high \pt electron or muon, or a lepton$+$jets trigger,
requiring a lower-\pt electron or muon accompanied by a jet. Events are required
to have at least one PV with $> 2$ tracks reconstructed within the fiducial
region of the SMT.  We require exactly four jets with $|\eta|<2.5$, with the
leading (highest \pt) jet having $\pt>40$ GeV, and the other jets $\pt>20$ GeV. 
Leptons are required to originate from within 1 cm of the PV in the coordinate
along the beam line. Exactly one tight electron (or tightly isolated muon) with
$\pt>20$ GeV and $|\eta|<1.1$ ($|\eta|<2$) is also required.  Electron$+$jets
events containing a second tight electron with $\pt>15$ GeV and $|\eta|<2.5$ or
a veto muon with  $\pt>15$ GeV and $|\eta|<2$ are rejected.  Muon$+$jets events
containing a second muon that is a veto muon with $\pt>15$ GeV and $|\eta|<2$ or
a tight electron with $\pt>15$ GeV and $|\eta|<2.5$ are rejected.  The missing
transverse momentum is required to satisfy $\ptmiss>20$ GeV ($\ptmiss>25$ GeV)
for $e+$jets  ($\mu+$jets) events.  Multijet background, typically arising from
mismeasurement of lepton or jet energies, is suppressed by requiring a minimal
azimuthal separation between the lepton direction and the \ptmiss vector with
$\Delta\phi(e,\ptmiss)>0.7\pi-0.045\cdot\ptmiss$ for electrons and
$\Delta\phi(\mu,\ptmiss)>2.1-0.035\cdot\ptmiss$ for muons, with \ptmiss in GeV and
$\Delta\phi(\ell,\ptmiss)=|\phi_{\ell}-\phi_{\ptmiss}|$. Any $\mu+$jets events
with an invariant masss, $m_{\mu\mu}$, of the isolated muon and a second muon
(with $\pt>15$ GeV and even lower quality requirements than a loosely isolated
muon) of $70<m_{\mu\mu}<110$ GeV are rejected in order to suppress
$Z(\rightarrow\mu\mu)+$jets events.  The data sample satisfying the above
criteria consists of $825$ $e+$jets and $737$ $\mu+$jets events.  We further
require at least one jet to be identified as a $b$ jet, which yields the final
data samples of 312 $e+$jets and 303 $\mu+$jets events.

\section{Monte Carlo Samples}
\label{sec:mcsamples}
The MC events used to model the \ttbar signal and the \wjets background needed
for the calibration of the measurement (described in Sec. \ref{sec:calib}) are
generated using \alpgen \citep{alpgen} to simulate the hard-scattering process
and \pythia \citep{pythia} to simulate hadronization and shower evolution.  The
MLM matching scheme \citep{mlm} is employed to avoid overlaps between components
of the event belonging to the hard process, implemented through a matrix
element, and parton evolution (showering) into jets.  The \wjets background
samples are divided into two categories: $(i)$ $W+lp$ and $(ii)$
$W+(c\overline{c},lp)$ and $W+(b\overline{b},lp)$, where $lp$ (light partons)
denotes $u$, $d$, $s$-quarks, or gluons.  Although the individual processes are
produced with \alpgen which is a LO generator, the relative contributions
between the two categories are determined using next-to-leading order (NLO)
calculations, with next-to-leading logarithmic (NLL) corrections based on the
\mcfm MC generator~\citep{mcfm}.  The MC samples used to derive jet transfer
functions that correlate jet energies with those of partons in the \ttbar events
(described in Sec. \ref{sec:tfjet}) are generated using \pythia to simulate both
the hard-scattering process and the subsequent hadronization and shower
evolution for the events.  All MC samples are generated with CTEQ6L1 parton
distribution functions (PDFs)~\citep{cteq}, and passed through a full
\geant-based \citep{geant} simulation of the D0 detector.  To simulate the
effects from additional \ppbar interactions, events with no trigger requirements
selected from random \ppbar crossings in the collider data having the same
instantaneous luminosity profile as the data are overlayed on the fully
simulated MC events. This is then followed by the same reconstruction and
analysis chain as applied to data.

\section{Method of Analysis}
\label{sec:me_method}
The top-quark mass is measured using all kinematic information with a likelihood
technique based on probability densities (PDs) constructed, for each individual
event, from matrix elements of the processes contributing to the observed final
state.  This analysis technique, referred to as the matrix element (ME)
method~\citep{run1nature}, is described below.

\subsection{Matrix Element Method}
\label{sec:me_method1}
If the processes contributing to an observed event do not interfere, the total
PD for observing a given event is the sum of all contributing probabilities for
that specific final state. Assuming that \ttbar and \wjets production are the
only two contributions, the PD for observing each event is given in terms of the
top-quark mass \mtop, the jet energy scale factor \kjes dividing the energies of
all jets, and the fractions of \ttbar signal ($f$) and of \wjets background
($1-f$) in the data by:
\begin{eqnarray}
P_{\rm evt} & = & A(x)[f P_{\rm sig}(x;m_{\rm t},\kjes)\nonumber\\
& &\hspace{30pt}+(1-f)P_{\rm bkg}(x;\kjes)],
\end{eqnarray}
where $x$ represents the measured jet and lepton energies and angles; $A(x)$,
which depends only on $x$, accounts for the geometrical acceptance and
efficiencies; and $P_{\rm sig }$ and $P_{\rm bkg }$ are the PDs for \ttbar and
\wjets production, respectively.  For events satisfying $P_{\rm bkg }\gg P_{\rm
sig }$, the relative contribution of $P_{\rm sig}$ to $P_{\rm evt }$ is
negligible and has minimal influence on the determination of \mtop.  Multijet
events satisfy this condition and can therefore be represented by  $P_{\rm bkg
}$, as the event kinematics are far closer to \wjets than to \ttbar production.

Due to the finite detector resolution and the hadronization process, the
measured set $x$ for the observed events will not, in general, be identical to
the corresponding set $y$ of the original final-state partons and the
relationship between $x$ and $y$ is described by a transfer function. In
addition, the initial partons carry momenta $q_{1}$ and $q_{2}$ in the colliding
$p$ and $\overline{p}$.  To account for this complication, $P_{\rm sig }$ and
$P_{\rm bkg }$ must be integrated over all parton states contributing to the
observed set $x$. This involves a convolution of the partonic differential cross
section $d\sigma(y)$ with the PDFs and a transfer function $W(x,y;\,\kjes)$ that
relates $x$ and $y$:
\begin{eqnarray}
P(x,\alpha) & = & \frac{1}{\sigma(\alpha)}\int{\displaystyle
\sum_{\textrm{flavors}}}d\sigma(y,\alpha)dq_{1}dq_{2}f(q_{1})f(q_{2})\nonumber\\
& &\hspace{53pt}\times\hspace{2pt}W(x,y;\,\kjes),
\end{eqnarray}
where $\alpha$ represents the parameters to be determined in the analysis, the
sum runs over all possible initial-state parton flavors, and $f(q_{i})$ are
CTEQ6L1 PDFs for finding a parton of a given flavor and longitudinal momentum
fraction $q_i$ in the $p$ or $\overline{p}$. Detector resolution is taken into
account in $W(x,y;\,\kjes)$, representing the probability density for the
measured set $x$ to have arisen from the partonic set $y$. Dividing by the total
observed cross section for the process, $\sigma(\alpha)$, ensures $P(x;\alpha)$
is normalized to unity.

\begin{figure}
\begin{centering}
\includegraphics[width=0.49\columnwidth]{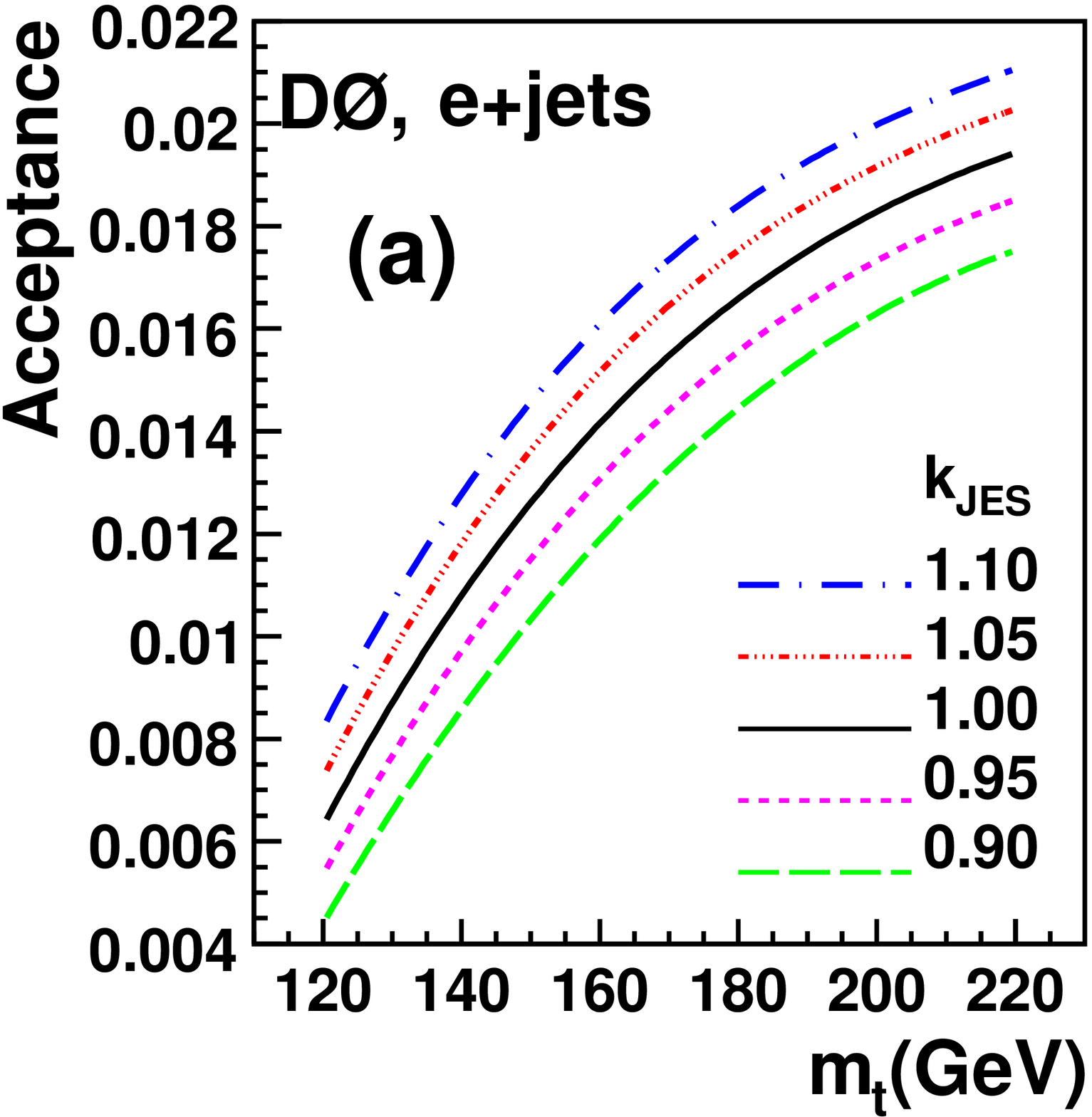}
\includegraphics[width=0.49\columnwidth]{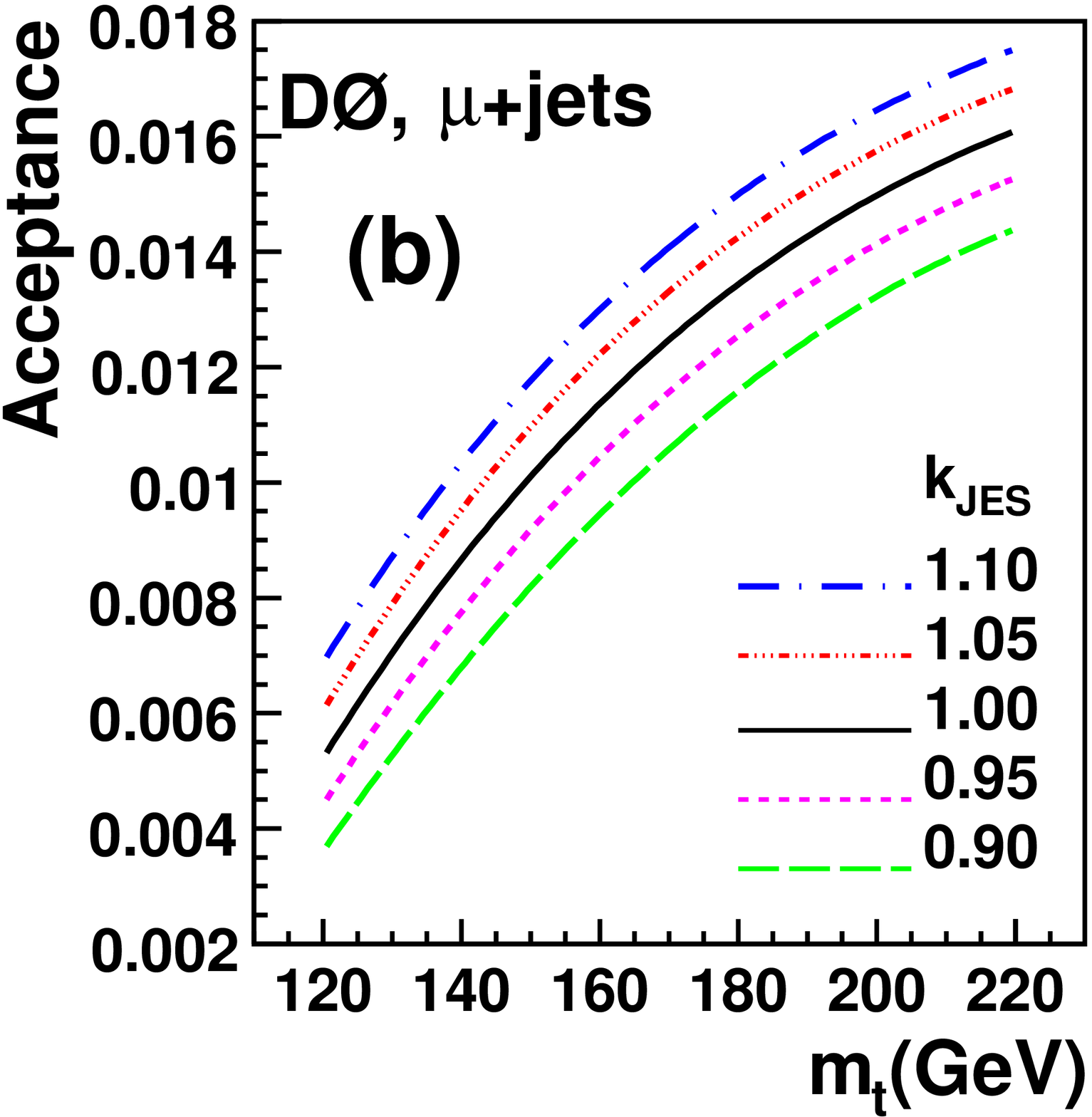}
\par\end{centering}
\caption{\label{fig:acc}(color online) The mean acceptance as a function of
\mtop and \kjes for the (a) $e$+jets and (b) $\mu$+jets channels.}
\end{figure}

The differential cross section term for $P_{\rm sig}$ is calculated using the LO
ME of the quark-antiquark annihilation process ($\mathcal{M}_{t\overline{t}})$.
A total of 24 integration variables are associated with the two initial state
partons and the six particles in the final state.  Since the angles for the four
jets and the charged lepton are sufficiently well measured, the angular
resolution terms in $W(x,y;\,\kjes)$ can be approximated by Dirac
$\delta$-functions. Integrating over these and four more $\delta$-functions that
impose energy-momentum conservation leaves 10 integrals to evaluate the
probability density that represents the $t\overline{t}$ production probability
for a given \mtop and
\kjes~\citep{integ}:
\begin{widetext}
\begin{eqnarray}
P_{\rm sig} & = & \frac{1}{\sigma_{\rm{obs}}^{t\overline{t}}}
{\displaystyle \sum_{i=1}^{24}}w_{i}{\displaystyle \int}
d\rho \hspace{1pt}dm_{1}^{2}\hspace{1pt} dM_{1}^{2}\hspace{1pt}dm_{2}^{2}\hspace{1pt}dM_{2}^{2}\hspace{1pt}
d\rho_{\ell}\hspace{2pt}dq_{1}^{\mathrm{x}}\hspace{1pt}dq_{1}^{\mathrm{y}}\hspace{1pt}dq_{2}^{\mathrm{x}}\hspace{1pt}dq_{2}^{\mathrm{y}}
{\displaystyle
\sum_{\textrm{flavors},\nu}}|\mathcal{M}_{t\overline{t}}|^{2}\frac{f^{'}(q_{1})f^{'}(q_{2})}{\sqrt{(\eta_{\alpha\beta}{q}_{1}^{\alpha}
{q}_{2}^{\beta})^{2}-m_{\rm q_{1}}^{2}m_{\rm q_{2}}^{2}}}\Phi_{6}W(x,y;\,\kjes),\hspace{10pt}\label{eq:psigint}
\end{eqnarray}
\end{widetext}
where, in addition to the CTEQ6L1 PDF given by $f(q_{i})$, the $f^{'}(q_{i})$
also include parameterizations of the probability distributions for the
transverse momenta $q_i^{\rm x,y}$ of the initial-state partons in
\pythia\citep{pythia}. The masses of the initial-state partons are given by
$m_{\rm q_{i}}$, and $\Phi_{6}$ includes the 6-body phase-space factor and other
constants. The first sum is over all 24 jet permutations, each carrying a weight
$w_{i}$, which is the product of four jet weights.  The weight for a $b$-tagged
jet with a given \pt and $\eta$ is the average tagging efficiency
$\epsilon_\alpha(\pt,\eta)$ for a given parton hypothesis $\alpha\,(=b,c,{\rm
light}\,q, {\rm or\,gluon})$.  The weight for a jet that is not $b$-tagged is
$1-\epsilon_\alpha(\pt,\eta)$. The second sum includes up to eight solutions for
neutrino kinematics, and conservation of transverse momentum used to calculate
the transverse momentum of the neutrino. The parameter $\rho$ represents the
fraction of the energy carried by one of the quarks from the $W\rightarrow
q\overline{q}^{\prime}$ decay. The masses of the two $W$ bosons
($M_{1}$,$M_{2}$) and of the pair of top quarks ($m_{1}$, $m_{2}$) are chosen as
integration variables because of computational efficiency related to the four
Breit-Wigner mass terms that make the ME negligible everywhere except at the
mass peaks. The energy (the curvature $1/p_{T}$) of the electron (muon) is
defined by $\rho_{\ell}$. The integration over $q_{i}$ involves only transverse
components. $W(x,y;\,\kjes)$ is the product of five terms for the four jets and
one charged lepton, described below. The normalization $\sigma_{\rm
obs}^{\ttbar}=\int{A(x)\Psig dx}=\sigma^{\ttbar}(\mtop)\times\langle
A(\mtop,\kjes)\rangle$ is calculated from the product of the total cross section
corresponding to the ME used and the mean acceptance for events whose
dependencies on \mtop and \kjes are determined from MC events.  The mean
acceptance is shown in Fig.~\ref{fig:acc} as a function of \mtop for different
values of \kjes.

The differential cross section in $P_{\textrm{bkg}}$ is calculated using
the $W+$4 jets matrix elements from the \vecbos \citep{vecbos} MC
program. The initial-state partons are assumed to have no transverse momenta. 
The integration is performed over the $W$ boson mass, the energy ($1/p_{T}$) of
the electron (muon), and the energies of the four partons producing the jets,
summing over the 24 jet permutations and all neutrino solutions.

The top-quark mass is extracted from $n$ events with a measured set of variables
$\xt=(x_{1},x_{2},...,x_{n})$ through a likelihood function for individual event
probabilities $P_{\rm evt}$ according to
\begin{equation}
L(\xt;\,\mtop,\kjes,f)=\prod_{i=1}^{n}P_{{\rm evt}}(x_{i};\,\mtop,\kjes,f).\label{eq:likejoint}
\end{equation}
For every assumed pair of $(\mtop,\ \kjes)$ values, the value of $f^{\rm best}$
that maximizes the likelihood is determined. To obtain the best estimate of
\mtop and \kjes, the two-dimensional likelihood:
\begin{equation}
L\left(\xt;\, \mtop,\kjes\right)=L\left[\xt;\, \mtop,\kjes,f^{\rm best}(\mtop,\kjes)\right]
\end{equation}
is projected onto the \mtop and \kjes axes according to
\begin{equation}
L\left(\xt;\, \mtop\right)=\int L\left(\xt;\,
\mtop,\kjes\right)G\left(\kjes\right)d\kjes\label{eq:likeproj_mtop}
\end{equation}
and
\begin{equation}
L\left(\xt;\, \kjes\right)=\int L\left(\xt;\,
\mtop,\kjes\right)d\mtop\label{eq:likeproj_jes},
\end{equation}
using Simpson's rule \citep{koonin}, where the prior probability distribution
$G\left(\kjes\right)$ is a Gaussian function centered at $\kjes=1$ with standard
deviation (sd) $0.02$ determined from the mean of the fractional uncertainty of
the standard jet energy scale corrections applied to all jets in the MC samples
used in this analysis.  The best estimates and the uncertainties on the mass
of the top quark and the jet energy scale are then extracted using the mean and
the RMS of $L\left(\xt;\,\mtop\right)$ and  $L\left(\xt;\, \kjes\right)$,
respectively. The mean is calculated from $\overline{\alpha} = \int{\alpha
L(\xt;\, \alpha)d\alpha}/\int{L(\xt;\, \alpha)d\alpha}$ and the RMS from
$\sigma^2(\alpha)=\int{(\alpha - \overline{\alpha})^2 L(\xt;\,
\alpha)d\alpha}/\int{L(\xt;\, \alpha)d\alpha}$, where $\alpha$ corresponds to
\mtop or \kjes, also using Simpson's rule.

The fit parameter \kjes, associated with the {\sl in situ} jet energy calibration,
has the effect of rescaling the energies of all the jets, and thereby the 2-jet
invariant mass of the hadronically decaying $W$ boson, with the jet energy scale
factor \kjes.  The presence of the Breit-Wigner mass term associated with the
hadronically decaying $W$ boson in the ME of Eq.~(\ref{eq:psigint}) maximizes
the likelihood in Eq.~(\ref{eq:likejoint}) when the 2-jet invariant mass
coincides with the Breit-Wigner pole fixed at the world average of $M_W=80.4$
GeV~\citep{pdg}.  The additional constraint to the standard scale derived from
$\gamma+$jet and dijet samples is applied through the prior probability
distribution $G\left(\kjes\right)$ in Eq.~(\ref{eq:likeproj_mtop}).
\begin{table}[htbp]
\caption{\label{tab:lj_tfpars}
Transfer function parameters for light quarks  ($a_i$ in GeV).}
\begin{center}
\begin{tabular}{lr@{$\times$}lr@{$\times$}lr@{$\times$}lr@{$\times$}l}
\hline
\hline
& \multicolumn{8}{c}{Light-quark jets} \\
Par. & \multicolumn{2}{c}{$|\eta|<0.5$} & \multicolumn{2}{c}{$0.5<|\eta|<1$} & \multicolumn{2}{c}{$1<|\eta|<1.5$} & \multicolumn{2}{c}{$1.5<|\eta|<2.5$} \\
\hline\vspace{-3mm}\\
$a_1$ &$-2.74$ & $10^{ 0}$ &$-8.02$ & $10^{-1}$ &$ 1.69$ & $10^{-1}$ &$ 1.52$ & $10^{ 1}$ \\
$b_1$ &$ 1.67$ & $10^{-2}$ &$-3.59$ & $10^{-3}$ &$ 1.32$ & $10^{ 1}$ &$-2.17$ & $10^{-1}$ \\
$a_2$ &$ 5.44$ & $10^{ 0}$ &$ 5.40$ & $10^{ 0}$ &$-3.26$ & $10^{-1}$ &$ 3.34$ & $10^{ 0}$ \\
$b_2$ &$ 6.29$ & $10^{-2}$ &$ 8.46$ & $10^{-2}$ &$ 6.97$ & $10^{ 0}$ &$ 1.45$ & $10^{-1}$ \\
$b_3$ &$ 4.30$ & $10^{-4}$ &$ 4.80$ & $10^{-4}$ &$ 2.52$ & $10^{-2}$ &$ 4.06$ & $10^{-3}$ \\
$a_4$ &$ 1.54$ & $10^{ 1}$ &$ 2.00$ & $10^{ 1}$ &$ 4.71$ & $10^{ 0}$ &$ 1.72$ & $10^{ 1}$ \\
$b_4$ &$-2.12$ & $10^{-1}$ &$-2.38$ & $10^{-1}$ &$-8.37$ & $10^{-3}$ &$-3.69$ & $10^{-2}$ \\
$a_5$ &$ 1.77$ & $10^{ 1}$ &$-2.38$ & $10^{-1}$ &$ 1.03$ & $10^{ 1}$ &$ 1.75$ & $10^{ 1}$ \\
$b_5$ &$ 1.96$ & $10^{-1}$ &$ 1.89$ & $10^{ 1}$ &$ 6.42$ & $10^{-2}$ &$ 5.34$ & $10^{-2}$ \\
\hline
\hline
\end{tabular}
\end{center}
\end{table}
\begin{table*}[htbp]
\caption{\label{tab:bj_tfpars}
Transfer function parameters for $b$-quark jets without and with a muon within
the jet cone ($a_i$ in GeV). }
\begin{center}
\begin{tabular}{lr@{$\times$}lr@{$\times$}lr@{$\times$}lr@{$\times$}l|lr@{$\times$}lr@{$\times$}lr@{$\times$}lr@{$\times$}l}
\hline
\hline
& \multicolumn{8}{c|}{$b$-quark jets without a muon within the jet cone} & &
\multicolumn{8}{c}{$b$-quark jets with a muon within the jet cone} \\
Par. & \multicolumn{2}{c}{$|\eta|<0.5$} & \multicolumn{2}{c}{$0.5<|\eta|<1$} & \multicolumn{2}{c}{$1<|\eta|<1.5$} & \multicolumn{2}{c|}{$1.5<|\eta|<2.5$} &
~Par. & \multicolumn{2}{c}{$|\eta|<0.5$} & \multicolumn{2}{c}{$0.5<|\eta|<1$} & \multicolumn{2}{c}{$1<|\eta|<1.5$} & \multicolumn{2}{c}{$1.5<|\eta|<2.5$} \\
\hline\vspace{-3mm}\\
$a_1$ &$ 3.30$ & $10^{ 0}$ &$ 5.38$ & $10^{ 0}$ &$ 2.85$ & $10^{ 0}$ &$ 1.38$ & $10^{ 1}$~  &  ~$a_1$ &$ 6.37$ & $10^{ 0}$ &$ 6.31$ & $10^{ 0}$ &$ 8.00$ & $10^{ 0}$ &$ 1.65$ & $10^{ 1}$ \\
$b_1$ &$-2.13$ & $10^{-1}$ &$-2.26$ & $10^{-1}$ &$-1.85$ & $10^{-1}$ &$-2.90$ & $10^{-1}$   &  ~$b_1$ &$-1.46$ & $10^{-1}$ &$-1.40$ & $10^{-1}$ &$-1.39$ & $10^{-1}$ &$-1.91$ & $10^{-1}$ \\
$a_2$ &$ 5.02$ & $10^{ 0}$ &$ 5.08$ & $10^{ 0}$ &$ 9.78$ & $10^{-1}$ &$ 3.86$ & $10^{ 0}$   &  ~$a_2$ &$ 2.53$ & $10^{ 0}$ &$ 3.89$ & $10^{ 0}$ &$ 8.54$ & $10^{ 0}$ &$ 4.88$ & $10^{ 0}$ \\
$b_2$ &$ 1.73$ & $10^{-1}$ &$ 1.77$ & $10^{-1}$ &$ 1.83$ & $10^{-1}$ &$ 1.36$ & $10^{-1}$   &  ~$b_2$ &$ 1.43$ & $10^{-1}$ &$ 1.37$ & $10^{-1}$ &$ 1.28$ & $10^{-1}$ &$ 1.43$ & $10^{-1}$ \\
$b_3$ &$ 3.48$ & $10^{-2}$ &$ 2.49$ & $10^{-2}$ &$ 6.69$ & $10^{-3}$ &$ 7.52$ & $10^{-3}$   &  ~$b_3$ &$ 3.90$ & $10^{-4}$ &$ 3.40$ & $10^{-4}$ &$ 1.90$ & $10^{-4}$ &$ 1.20$ & $10^{-4}$ \\
$a_4$ &$-6.68$ & $10^{ 0}$ &$-6.56$ & $10^{ 0}$ &$ 8.54$ & $10^{-1}$ &$ 5.59$ & $10^{ 0}$   &  ~$a_4$ &$ 2.80$ & $10^{ 1}$ &$ 1.52$ & $10^{ 1}$ &$ 7.89$ & $10^{ 1}$ &$ 4.73$ & $10^{ 1}$ \\
$b_4$ &$ 2.38$ & $10^{-2}$ &$ 1.91$ & $10^{-2}$ &$-2.83$ & $10^{-2}$ &$-4.54$ & $10^{-2}$   &  ~$b_4$ &$-3.87$ & $10^{-1}$ &$-9.74$ & $10^{-2}$ &$ 2.22$ & $10^{-1}$ &$ 5.21$ & $10^{-2}$ \\
$a_5$ &$ 5.06$ & $10^{ 0}$ &$ 4.36$ & $10^{ 0}$ &$ 1.38$ & $10^{ 1}$ &$ 1.50$ & $10^{ 1}$   &  ~$a_5$ &$ 1.80$ & $10^{ 1}$ &$ 2.32$ & $10^{ 1}$ &$ 2.80$ & $10^{ 1}$ &$ 2.83$ & $10^{ 1}$ \\
$b_5$ &$ 4.71$ & $10^{-2}$ &$ 6.99$ & $10^{-2}$ &$ 6.04$ & $10^{-2}$ &$ 7.60$ & $10^{-2}$   &  ~$b_5$ &$ 1.30$ & $10^{-1}$ &$ 2.91$ & $10^{-2}$ &$-2.87$ & $10^{-1}$ &$-8.55$ & $10^{-2}$ \\
\hline
\hline
\end{tabular}
\end{center}
\end{table*}
\subsection{Detector Resolution}
In this section, we describe the parameterizations for the jet and electron
energy and muon \pt resolutions used in the transfer function $W(x,y;\,\kjes)$
which is the product of four jet transfer functions for a given jet permutation
and an electron or muon transfer function.

\subsubsection{Parameterization of Jet Energy Resolution}
\label{sec:tfjet}
\begin{figure}
\begin{centering}
\includegraphics[width=0.49\columnwidth]{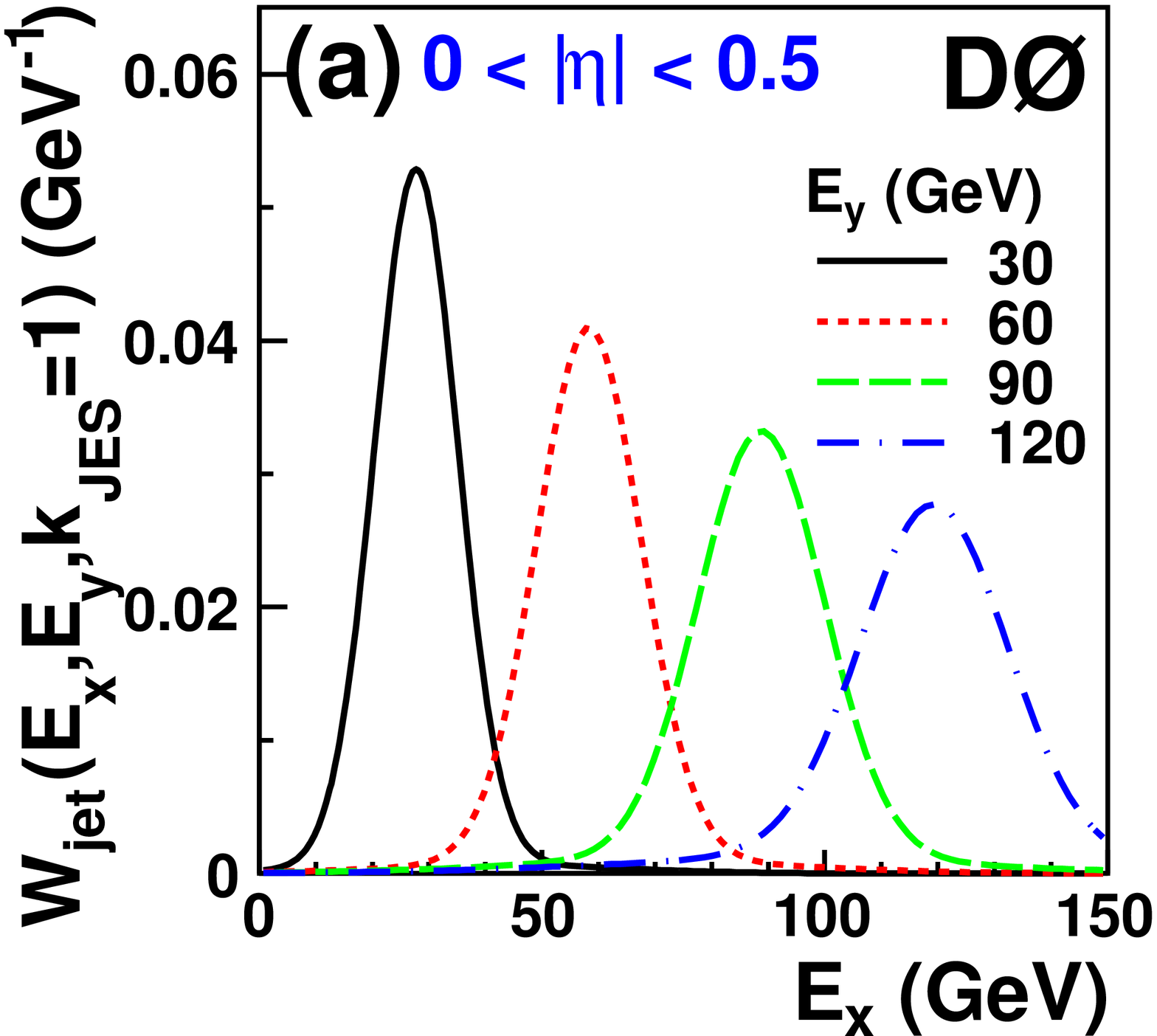}\includegraphics[width=0.49\columnwidth]{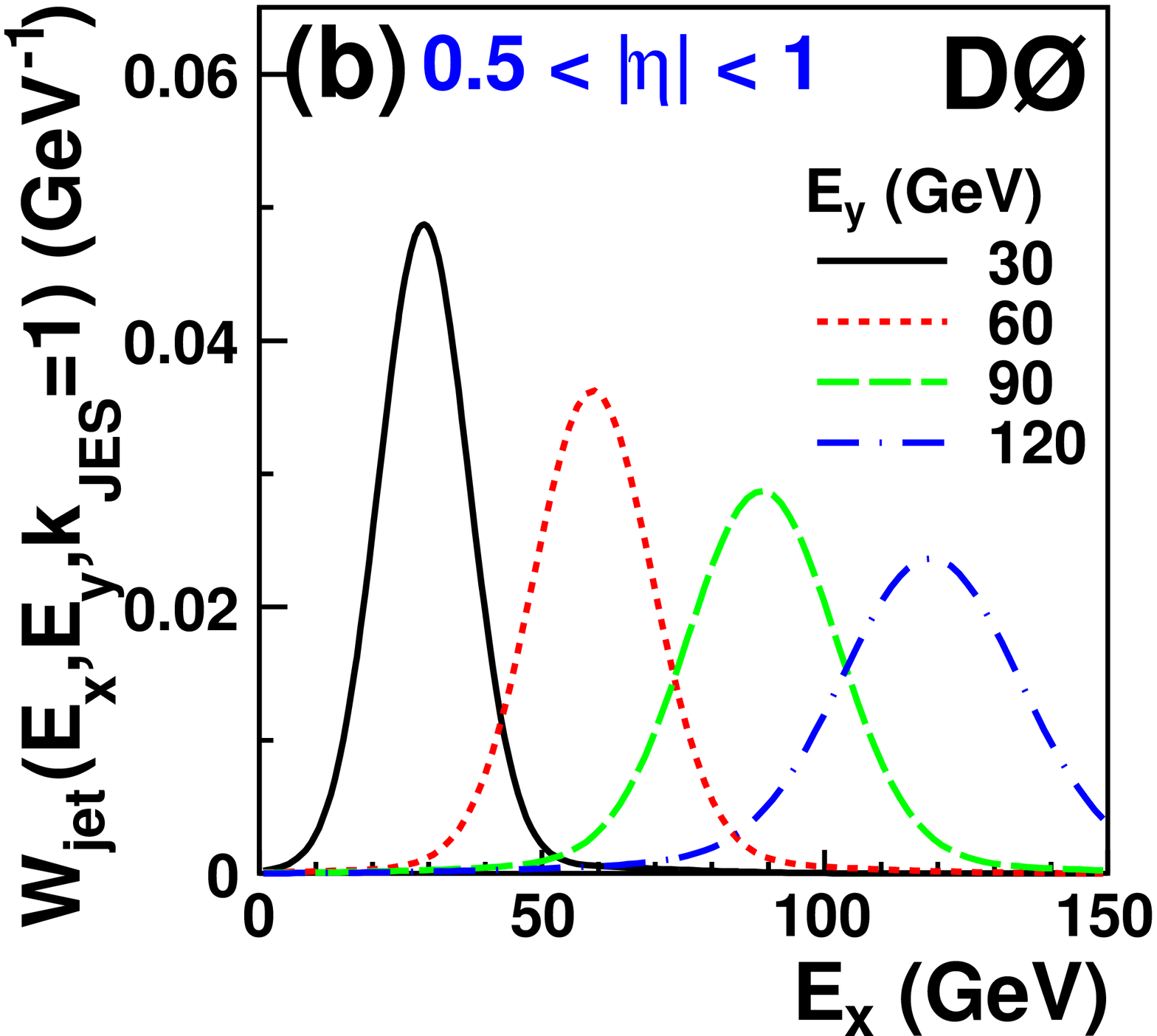}
\includegraphics[width=0.49\columnwidth]{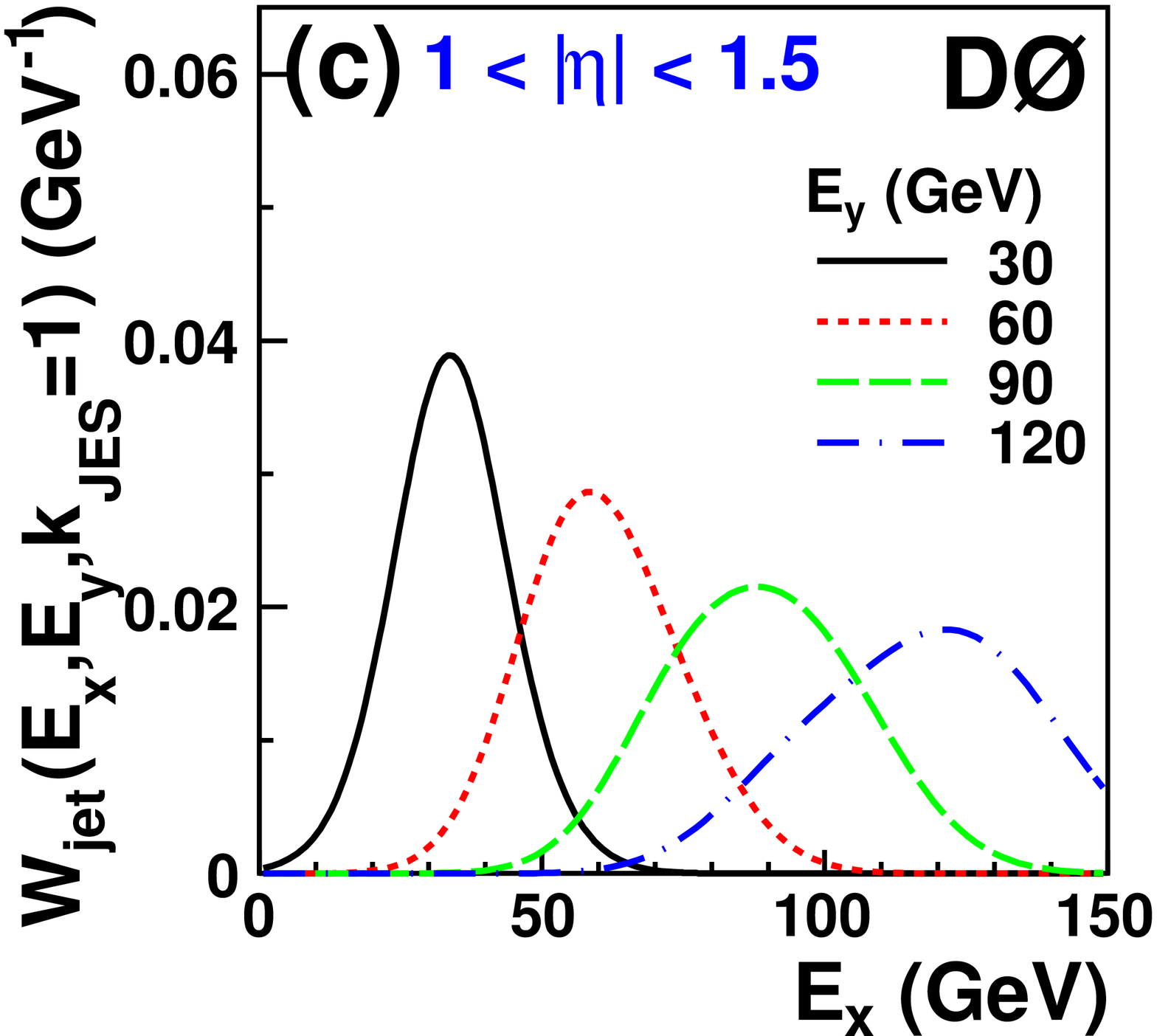}\includegraphics[width=0.49\columnwidth]{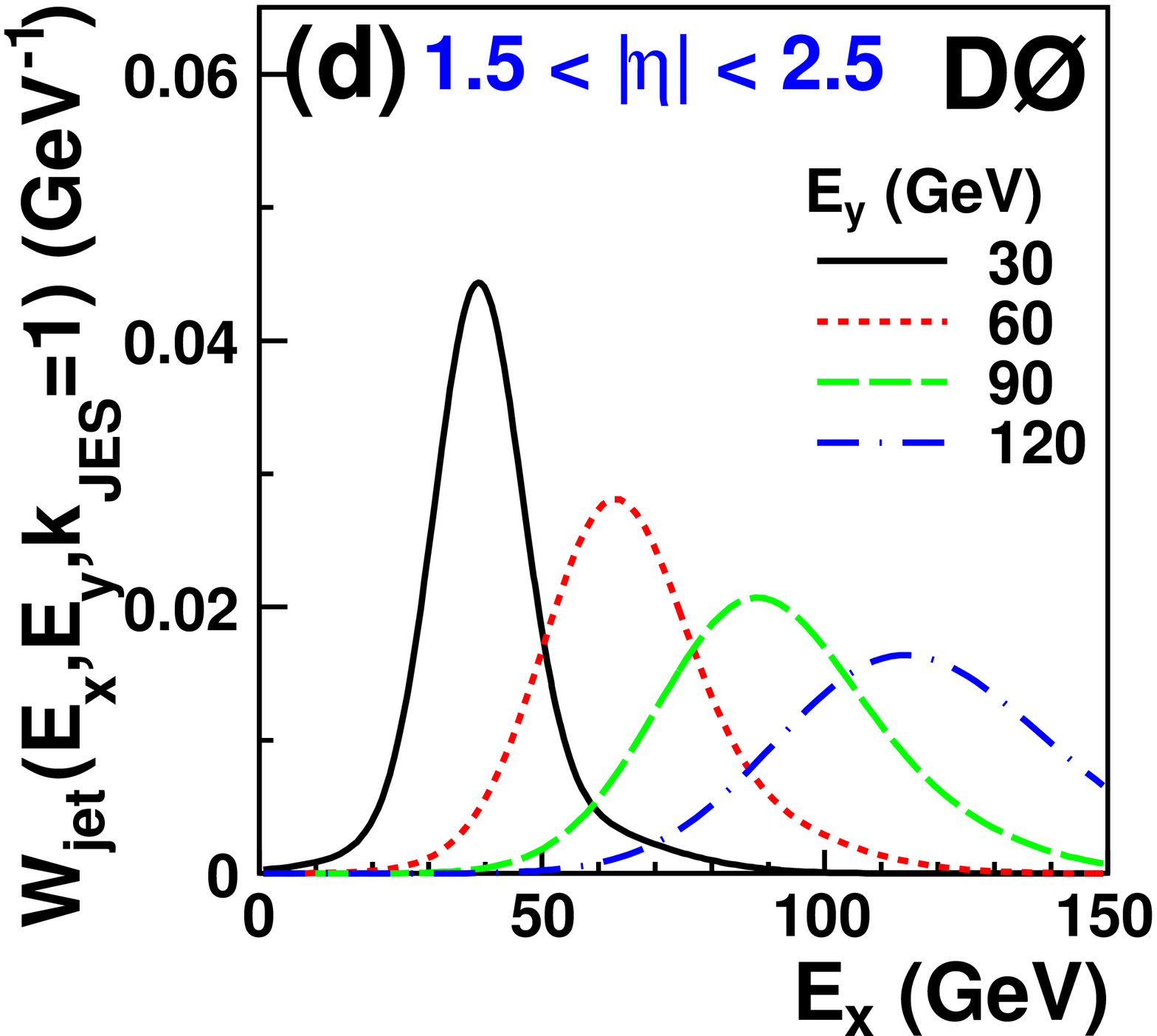}
\par\end{centering}
\caption{\label{fig:lj_tf}(color online) Transfer functions for \kjes=1 light-quark jets as a
function of measured jet energy for different parton energies in $\eta$ regions:
(a) $|\eta|<0.5$, (b) $0.5<|\eta|<1.0$, (c) $1.0<|\eta|<1.5$, and (d)
$1.5<|\eta|<2.5$.}
\end{figure}

\begin{figure}
\begin{centering}
\includegraphics[width=0.49\columnwidth]{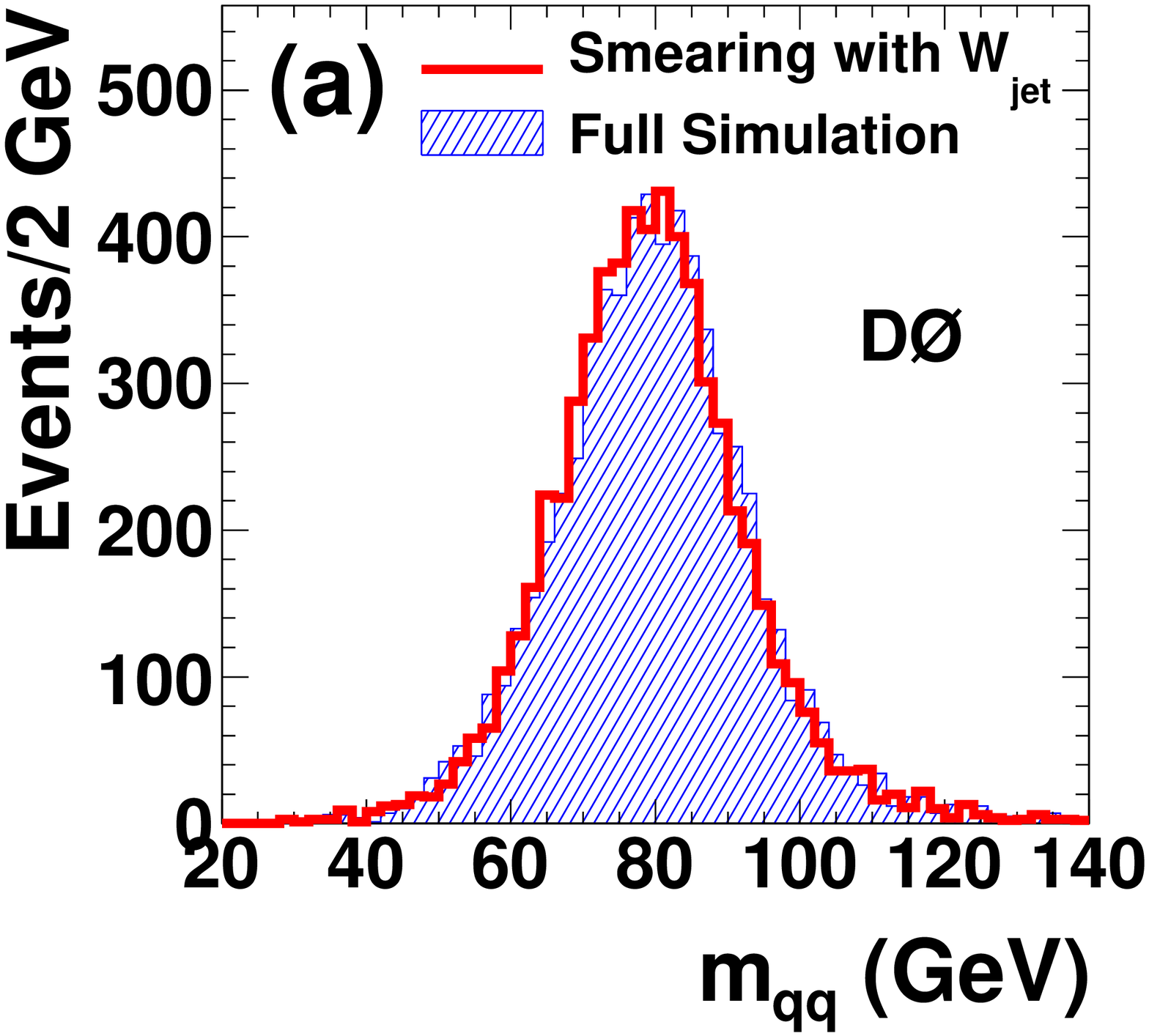}
\includegraphics[width=0.49\columnwidth]{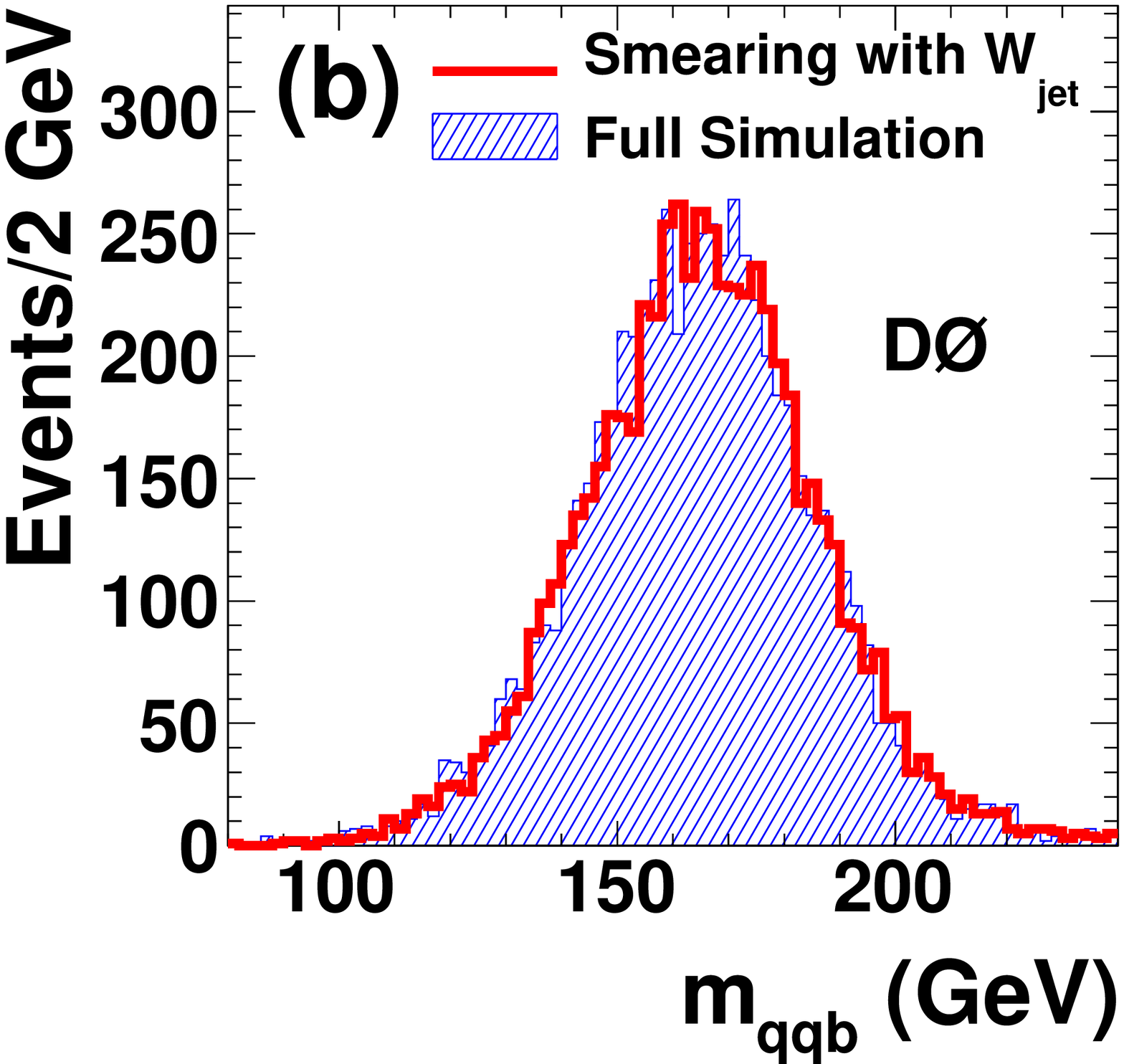}
\par\end{centering}
\caption{\label{fig:tfcheck}(color online) Comparison of (a) 2-jet and (b) 3-jet invariant mass
distributions for parton-level \ttbar MC events with energies smeared using the
transfer functions (open histogram) and fully simulated \ttbar MC events with
all four jets spatially matched to partons (filled histogram).}
\end{figure}

The transfer function for jets, $W_{{\rm jet}}(E_{x},E_{y};\kjes)$, represents
the probability that a measured jet energy $E_{x}$ in the detector corresponds
to a parent quark of energy $E_{y}$. It is parameterized in terms of a
double Gaussian function whose means and widths are dependent on $E_y$.  For the
case $\kjes=1$, it is given by
\begin{eqnarray}
W_{{\rm jet}}\left(E_{x},E_{y};\,\kjes=1\right)&=&\frac{1}{\sqrt{2\pi}(p_{2}+p_{3}p_{5})}\nonumber\\
& &\hspace{1pt}\times\left[e^{-\frac{[(E_{x}-E_{y})-p_{1}]^{2}}{2p_{2}^{2}}}\right.\nonumber\\
& &\hspace{20pt}\left.+\,p_{3}e^{-\frac{[(E_{x}-E_{y})-p_{4}]^{2}}{2p_{5}^{2}}}\right],\hspace{10pt}
\end{eqnarray}
where the $p_{i}$ are functions of the quark energy for quark $i$ and are
parameterized as linear functions of the $E_y$:
\begin{equation}
p_{i}=a_{i}+E_{y}\cdot b_{i}.
\end{equation}
The parameters $a_{i}$ and $b_{i}$ are determined from fully simulated \ttbar
events, following all jet energy corrections and smearing to match resolutions
in data. These events are generated with \pythia at nine values of the top-quark
mass ranging from 155 to 195 GeV in 5 GeV intervals. The parton and jet energies
are used in an unbinned likelihood fit that minimizes the product of the $W_{\rm
jet}$ terms for each event with respect to $a_{i}$ and $b_{i}$. A different set
of parameters is derived (i) for three varieties of quarks: light quarks ($u$,
$d$, $s$, $c$), $b$ quarks with a soft muon tag in the jet~\citep{slt}, and all
other $b$ quarks, and (ii) for four $\eta$ regions: $|\eta|<0.5$,
$0.5<|\eta|<1.0$, $1.0<|\eta|<1.5$, and $1.5<|\eta|<2.5$, to minimize possible
effects due to non-uniform calorimeter response. The values for these parameters
are shown in Tables \ref{tab:lj_tfpars} and \ref{tab:bj_tfpars} for light-quark
and $b$-quark jets, respectively. Figure \ref{fig:lj_tf} illustrates the
transfer functions for light-quark jets as a function of $E_{x}$ for different
values of $E_{y}$. In Fig.~\ref{fig:tfcheck}, we compare the 2-jet and 3-jet
invariant mass distributions for two types of \pythia \ttbar \ljets events: (i)
parton level events with jet energies smeared using the transfer functions and
(ii) fully simulated events where all four reconstructed jets are matched to
partons with $\Delta {\cal R}({\rm parton},{\rm jet})<0.5$.  The 2-jet (3-jet)
invariant masses are calculated using the two light-quark jets (all three jets)
from the hadronic branch of the \ttbar \ljets events and correspond to the $W$
boson (top-quark) mass.  The overlaid distributions in Fig. \ref{fig:tfcheck}
indicate that the jet transfer functions describe the jet resolutions well.

For $\kjes\neq1$, the jet transfer function is changed to
\begin{equation}
W_{{\rm jet}}\left(E_{x},E_{y};\,\kjes\right)=\frac{W_{{\rm jet}}\left(\frac{E_{x}}{\kjes},E_{y};\,1\right)}{\kjes},
\end{equation}
where the \kjes factor in the denominator preserves the
normalization $\int{W_{\rm jet}(E_x,E_y;\kjes)dE_x=1}$.

\subsubsection{Parameterization of Energy Resolution for Electrons}

The electron energy resolution is parameterized by the transfer function
\begin{equation}
W_{e}\left(E_{x},E_{y}\right)=\frac{1}{\sqrt{2\pi}\sigma}\exp\left[-\frac{1}{2}\left(\frac{E_{x}-E_{y}^\prime}{\sigma}\right)^{2}\right],
\end{equation}
where $E_x$ is the reconstructed electron energy,
\begin{eqnarray}
E_{y}^\prime & = & 1.000\cdot E_y+0.324\,{\rm GeV},\\
\sigma & = & \sqrt{(0.028\cdot E_{y}^\prime)^{2}+(S\cdot E_{y}^\prime)^{2}+(0.4\,{\rm GeV})^{2}},\\
S & = & \frac{0.164\,{\rm GeV}^{\nicefrac{1}{2}}}{\sqrt{E_y^\prime}}+\frac{0.122\,{\rm GeV}^{\nicefrac{1}{2}}}{E_y^\prime}
e^{\nicefrac{C}{\sin\theta_e}}-C,\hspace{10pt}\\
C & = & 1.3519-\frac{2.0956\,{\rm GeV}}{E_{y}^\prime}-\frac{6.9858\,{\rm GeV}}{E_{y}^{\prime2}},
\end{eqnarray}
$E_y$ is the energy of the original electron, and $\theta_e$ is the polar angle
of the electron with respect to the proton beam direction.  The parameters above
are derived from the detailed modeling of electron energy response and
resolution used in Ref.~\citep{wmassprl}.

\subsubsection{Parameterization of Momentum Resolution for Muons}
We describe the resolution of the central tracker through the uncertainty on 
the signed curvature of a track, the ratio of the electric charge and of the
transverse momentum of a particle, parameterized as a function of
pseudorapidity. The muon transfer function is parameterized as
\begin{equation}
W_{\mu}\left(\kappa_x,\kappa_y\right)=\frac{1}{\sqrt{2\pi}\sigma}
\exp\left[-\frac{1}{2}\left(\frac{\kappa_{x}-\kappa_{y}}{\sigma}\right)^{2}\right],
\end{equation}
where $\kappa_x=\left(q/p_T\right)_x$ and $\kappa_y=\left(q/p_T\right)_y$, with
the charge $q$ and transverse momentum $p_{T}$ of the original muon ($y$) or its
reconstructed track ($x$). The resolution
\begin{equation}
\sigma = 
\left\{\begin{array}{cc}
\sigmat & {\rm for}\ |\eta|\le1.4\\
\sqrt{\sigmat^{2}+\left\{c\cdot\left(|\eta|-1.4\right)\right\}^{2}} & {\rm for}\
|\eta|>1.4
\end{array}\right.
\end{equation}
is obtained from muon tracks in simulated events where the \sigmat and $c$
parameters are linear functions of $1/\pt$:
\begin{eqnarray}
\sigmat & = & \sigmat_0+\sigmat_1\cdot1/\pt,\\
c & = & c_0+c_1\cdot1/\pt.
\end{eqnarray}

The values of the coefficients are given in Table \ref{tab:mu_tfpar} for muon
tracks with associated and no associated hits in the silicon tracker.  This
simplified parameterization of the momentum resolution is valid at high
transverse momenta ($p_{T}>20$~GeV) where the limitations in coordinate
resolution dominate over the effects of multiple scattering.

\begin{table}
\caption{Parameters for muon transfer functions for muon tracks with and without
hits in the SMT.}
\label{tab:mu_tfpar} 
\centering
\begin{tabular}{lcc}
\hline 
\hline 
\multirow{2}{*}{Parameter~}
          & With hits & No hits \\
          & in the SMT & in the SMT \\
\hline\vspace{-3mm}\\
$\sigmat_0 ~({\rm GeV}^{-1})$ &~$2.082\times10^{-3}$~&~$3.620\times10^{-3}$ \\
$\sigmat_1$             & $1.125\times10^{-2}$ & $1.388\times10^{-2}$ \\
$c_0~({\rm GeV}^{-1})$       & $7.668\times10^{-3}$ & $2.070\times10^{-2}$ \\
$c_1$                   & $7.851\times10^{-2}$ & $7.042\times10^{-2}$ \\
\hline
\hline 
\end{tabular}
\end{table}

\section{Calibration of the Measurement}
\label{sec:calib}

\begin{figure}
\begin{centering}
\includegraphics[width=0.49\columnwidth]{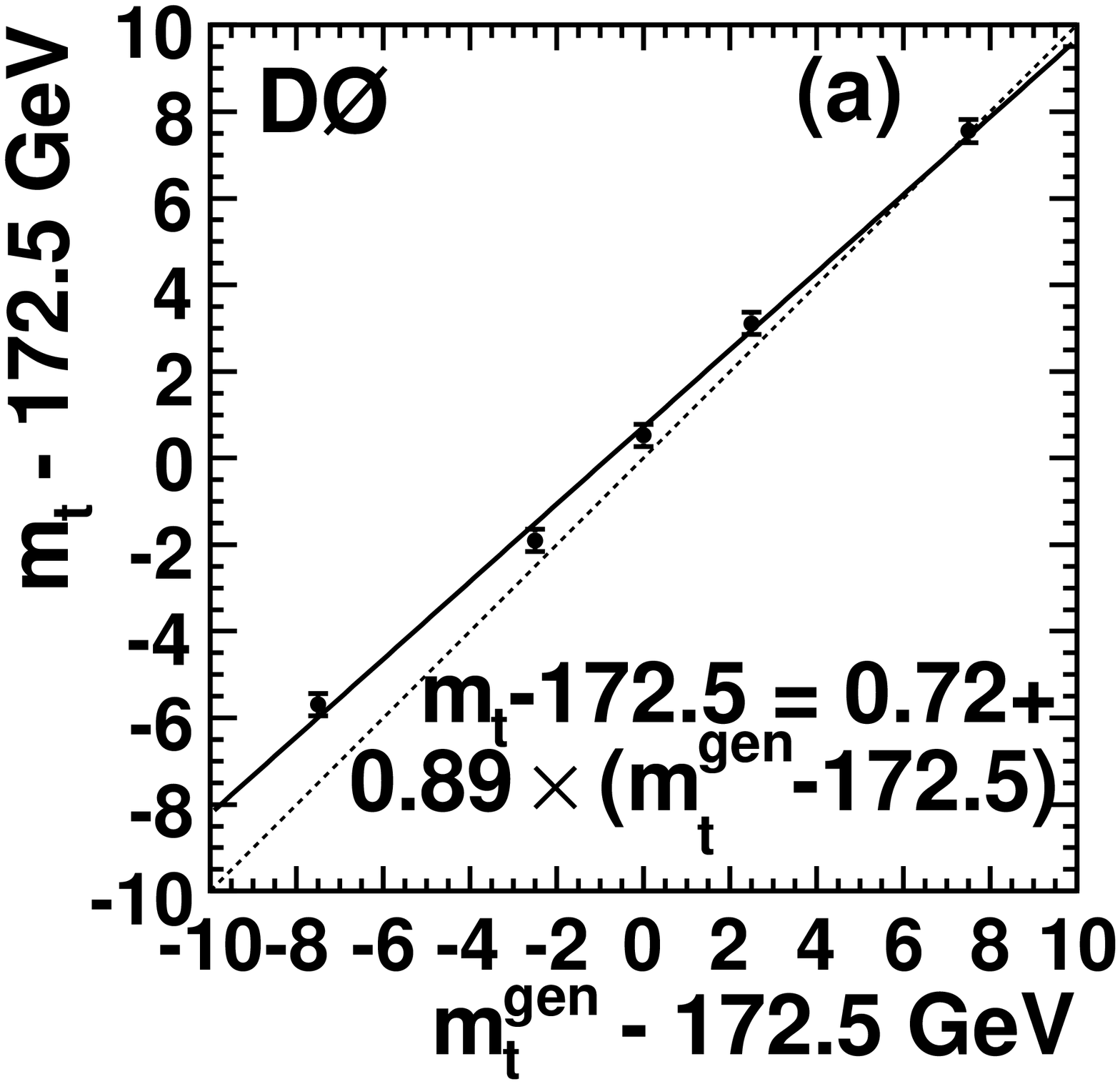}
\includegraphics[width=0.49\columnwidth]{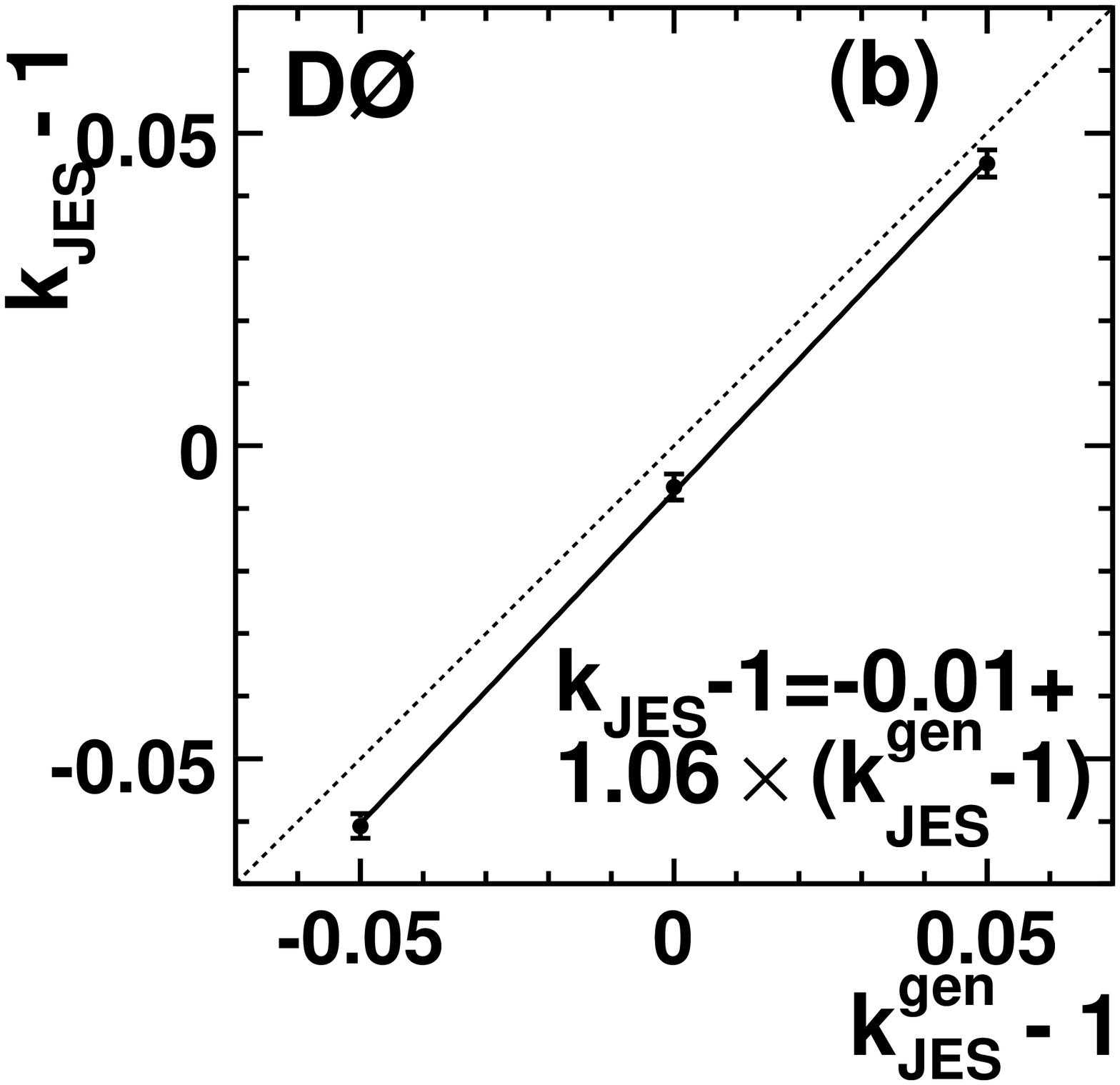}
\par\end{centering}
\caption{\label{fig:calib}Mean values of (a) \mtop and (b) \kjes extracted from
ensemble studies, as a function of the input values fitted to straight lines. 
Dashed lines represent 1:1 correlations of extracted and input values.}
\end{figure}
\begin{figure}
\begin{centering}
\includegraphics[width=0.49\columnwidth]{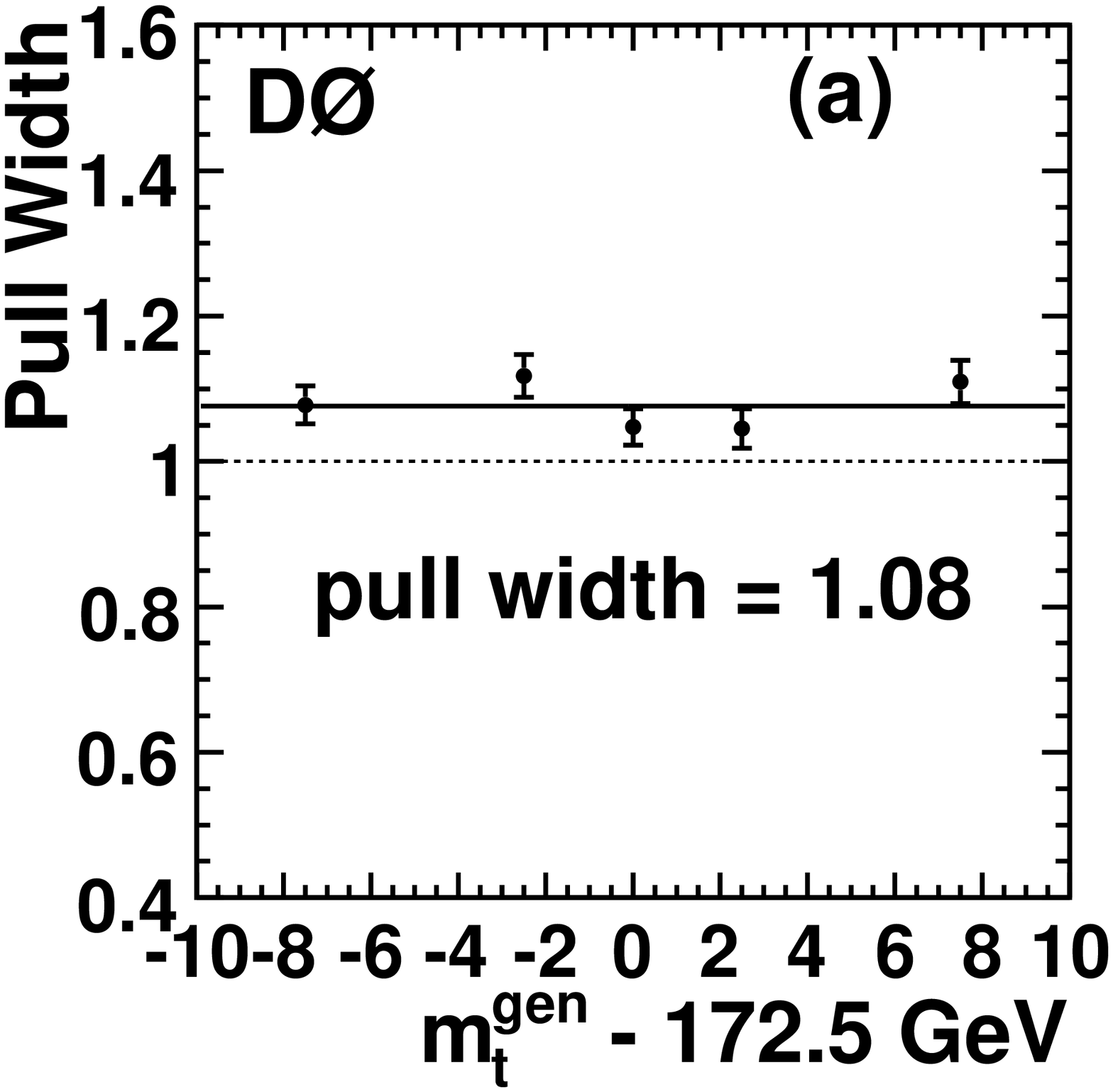}
\includegraphics[width=0.49\columnwidth]{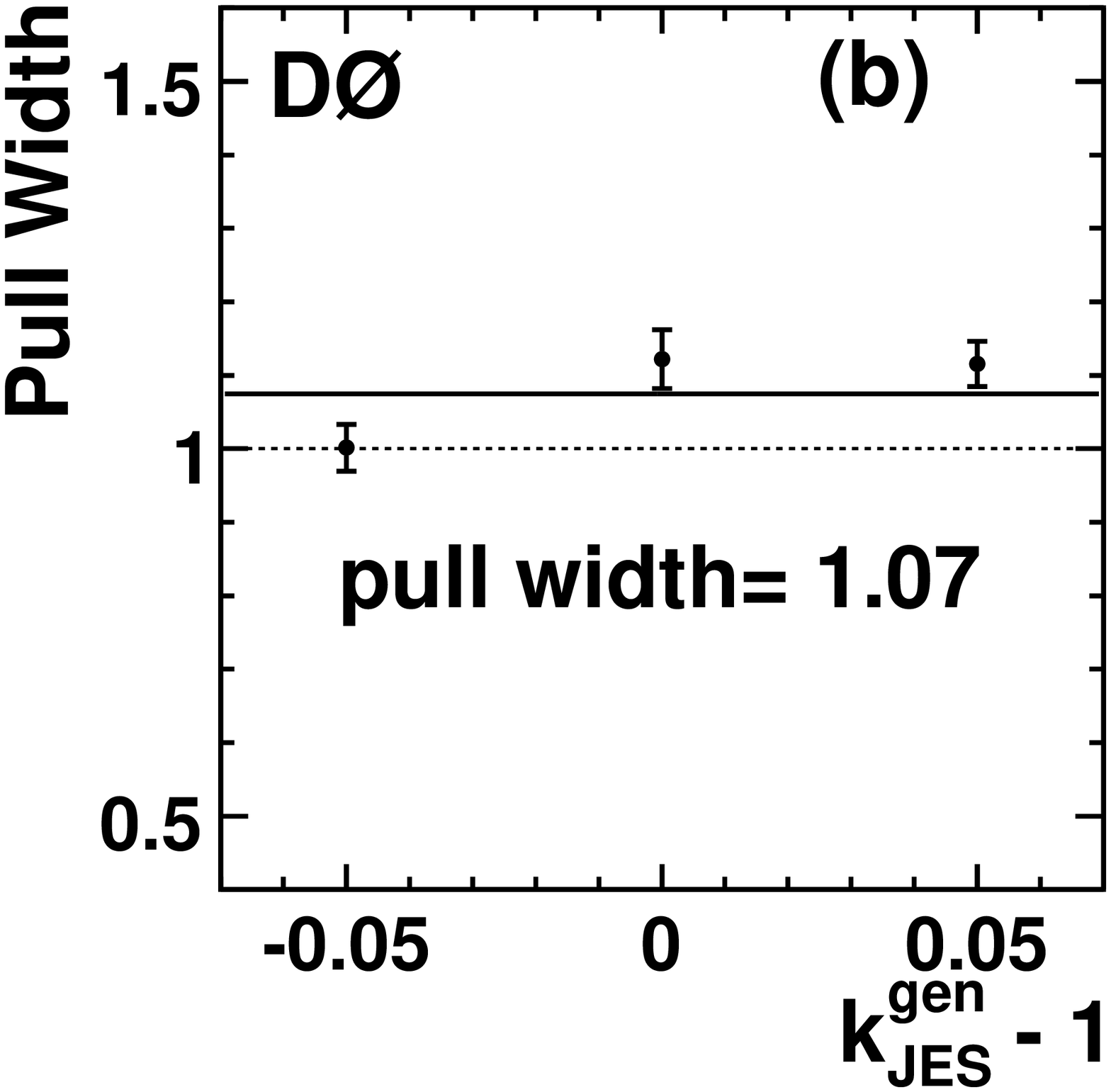}
\par\end{centering}
\caption{\label{fig:pulls}Widths of the pull distributions for (a) \mtop and (b)
\kjes from ensemble studies as a function of the input values.}
\end{figure}
The fully simulated MC samples described in Sec. \ref{sec:mcsamples} are used in
ensemble studies to calibrate the result from the ME method by determining and
correcting for biases in the extracted parameters and their estimated
uncertainties.  Such biases can be due, for example, to limitations in the LO ME
used in Eq.~(\ref{eq:psigint}) or to the imperfect description of detector
resolution using transfer functions with a limited number of parameters. Five
\ttbar MC samples are generated for $\mtgen=165$, $170$, $172.5$, $175$, and
$180$ GeV, with two more produced from the $172.5$ GeV sample by re-scaling all
jet energies by $\pm5$\%. \Psig and \Pbkg are calculated for these samples and
for the \wjets MC samples.  Events are drawn randomly from a \ttbar sample with
a particular mass and the \wjets sample to form pseudoexperiments, each with a
number of events equal to the one observed in data (before requiring $\geq 1$
$b$-tagged jets), with the signal fraction fluctuated according to a binomial
distribution relative to that determined from data. The values of \mtop and
\kjes are extracted for each pseudoexperiment according to the procedure
described in Sec.~\ref{sec:me_method1} using only events with at least one
$b$-tagged jet.  A thousand pseudoexperiments are performed for each of the 7
\ttbar samples.  The means (and their uncertainties) of all 1000 measured values
of \mtop and \kjes in each sample are determined from Gaussian fits to their
distributions and plotted versus the input $\mtgen - 172.5$ GeV and $\kjes - 1$,
respectively.  A straight line is fitted to the plotted points, representing the
response function used to correct the measurement from data (Fig.
\ref{fig:calib}).  For each pseudoexperiment, we also calculate the pulls,
defined as $(\mtop -
\langle\mtop\rangle)/\sigma(\mtop)$ and $(\kjes -
\langle\kjes\rangle)/\sigma(\kjes)$, where  $\langle\mtop\rangle$ and
$\langle\kjes\rangle$ are the mean measured \mtop and \kjes, respectively, for
all pseudoexperiments, and $\sigma(\mtop)$ and $\sigma(\kjes)$ are the RMS of
\mtop and \kjes, respectively, for the given pseudoexperiment.  The width of the
pull distributions for \mtop and \kjes are shown as a function of \mtgen and
\kjgen in Fig. \ref{fig:pulls}.  The average widths of the \mtop and \kjes pull
distributions are 1.08 and 1.07, respectively.

\begin{figure}
\begin{centering}
\includegraphics[width=0.49\columnwidth]{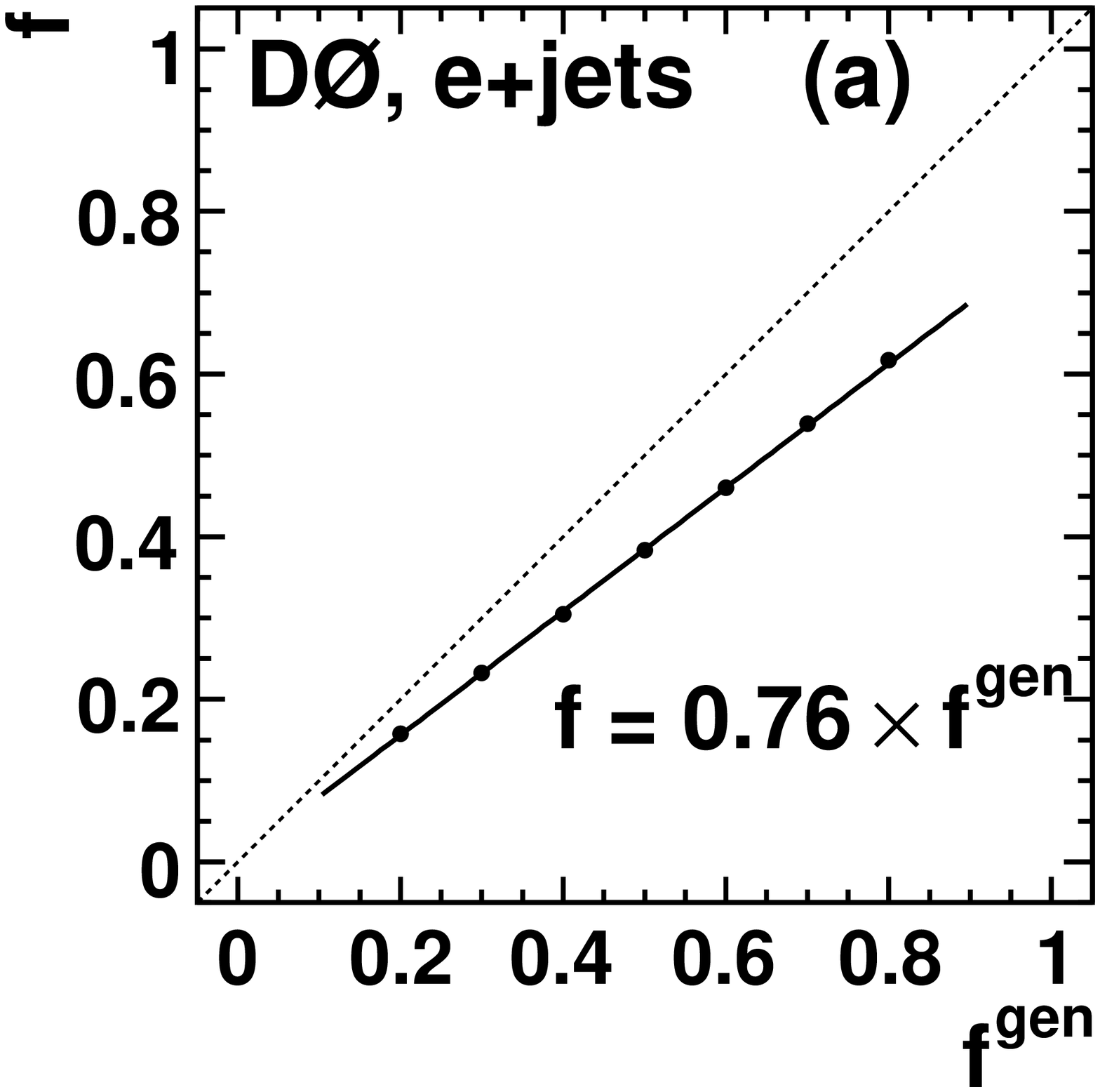}
\includegraphics[width=0.49\columnwidth]{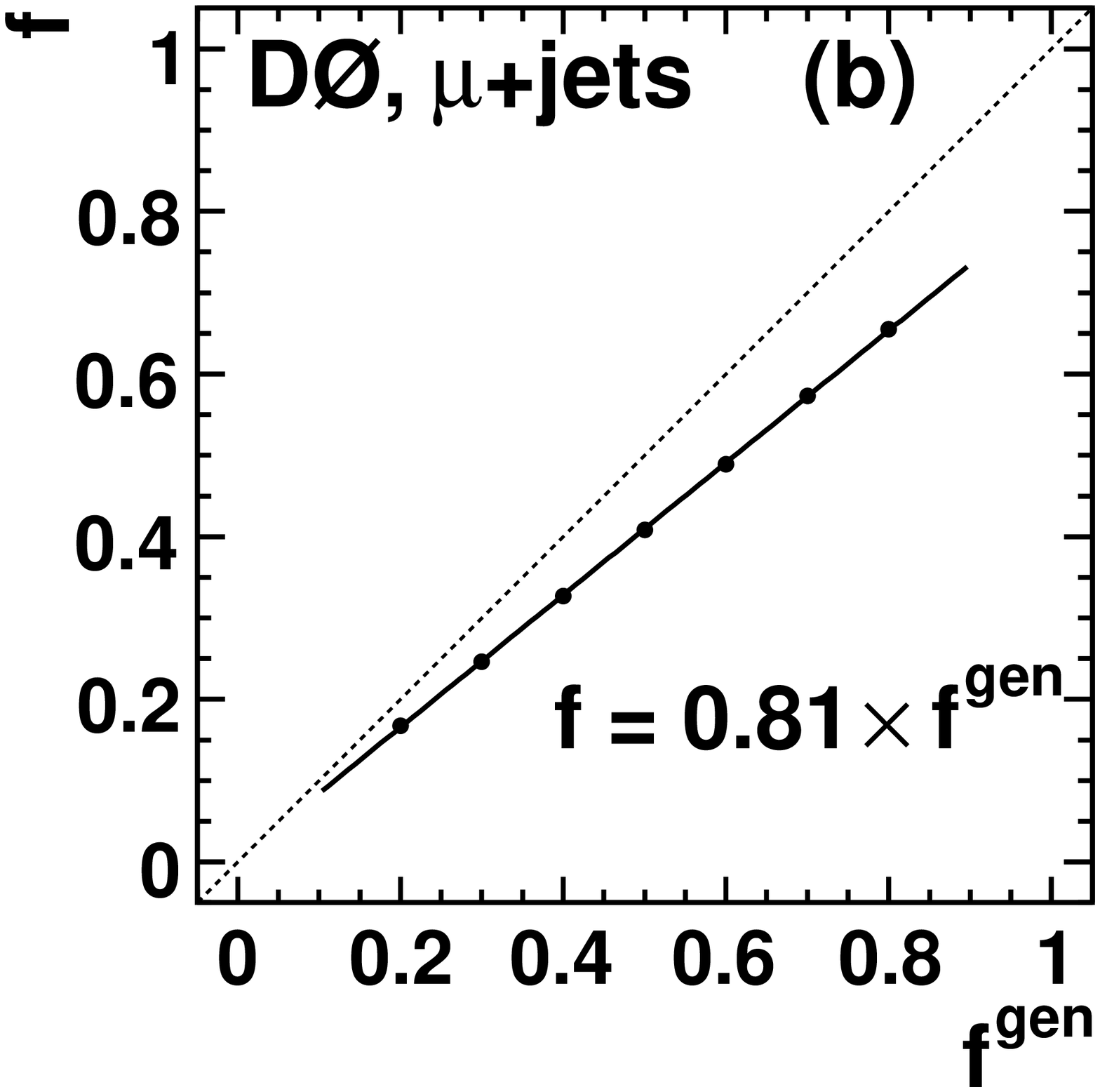}
\par\end{centering}
\caption{\label{fig:sfrac}Extracted signal fractions from ensemble studies as a
function of the input values for the (a) $e+$jets and (b) $\mu+$jets channels.}
\end{figure}

The signal fraction for the ensemble studies is determined from the selected
data sample using the method described in Sec.~\ref{sec:me_method1}.  To correct
for biases in the determination of this fraction, a calibration is done using
the \wjets and $172.5$ GeV \ttbar MC samples, wherein 1000 pseudoexperiments are
performed using the same procedure as described in the previous paragraph, but
with signal fractions set to a different value in each test.  The extracted
signal fractions as a function of their input values are shown in
Figs.~\ref{fig:sfrac}(a) and \ref{fig:sfrac}(b) for the $e+$jets and $\mu+$jets
channels, respectively.  Straight lines are fitted to the points in  plots
representing the response functions used to correct the fractions determined
from the selected data sample.  The calibration of the signal fraction is
performed separately for the $e+$jets and $\mu+$jets channels.  The corrected
fractions are $0.35\pm0.05$ and $0.41\pm0.06$ for the $e+$jets and $\mu+$jets
channels, respectively, prior to requiring at least one $b$-tagged jet.  These
fractions are $0.71\pm0.05$ and $0.75\pm0.04$ for the $e+$jets and $\mu+$jets
channels, respectively, after requiring at least one $b$-tagged jet.

\section{Flavor-Dependent Jet Response Correction for MC Events}
\label{sec:flavcorr}
\begin{turnpage}
\begin{figure*}
\begin{centering}
\includegraphics[width=1.3\textwidth]{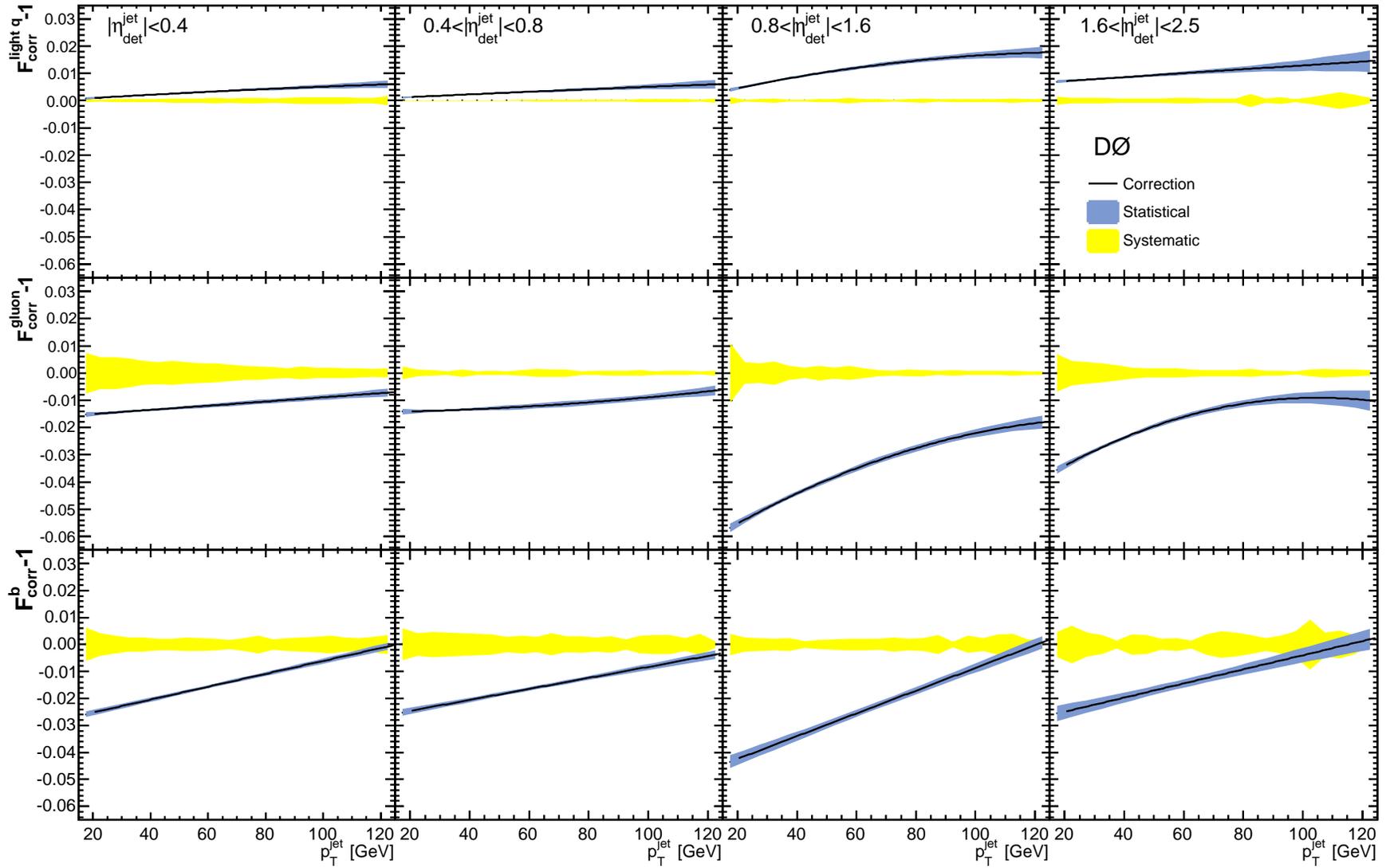}
\par\end{centering}
\caption{\label{fig:flavcorr}(color online) Correction factors for data-MC jet response
difference for light-quark, gluon, and $b$ jets as a function of jet \pt and
$\eta$. Light shaded bands represent statistical uncertainties and dark shaded
bands represent systematic uncertainties}
\end{figure*}
\end{turnpage}
The validity of the  calibration procedure described in the previous section is
based on the assumption of a perfect MC simulation of the events and of the
detector.  Uncertainties in this assumption are discussed in the section on
systematic uncertainties (Sec. \ref{sec:syst}).  The {\sl in situ} jet energy scale
employed in this analysis can account for a global scale discrepancy between
data and MC jet energy scales (see Sec. \ref{sec:rjes}) by rescaling the
energies of the two light jets from the $W\rightarrow q\overline{q}^{\prime}$
decay of \ttbar events to the world average mass of the $W$ boson~\citep{pdg}. 
This same rescaling is also applied to the two $b$ jets in the event.  However,
jets originating from different partons have different kinematic characteristics
and particle compositions.  In particular, $b$ and light jets with different
electromagnetic fractions can lead to different responses in a non-compensating
calorimeter.  Such features, if not properly simulated, can result in a
systematic shift in the determination of the top-quark mass.  In fact, the
largest contribution to the total systematic uncertainty of our previous
analysis in Ref.~\citep{meljprl} is the $b$/light-quark response ratio which was
an estimate of the effect of such a discrepancy.

To bring the simulation of the calorimeter response to jets into agreement with
data, and thereby reduce the systematic uncertainty associated with a jet
response difference in data and MC, we determine a flavor-dependent correction
factor as follows.  We note a discrepancy in the predicted energy deposition in
the calorimeter between data and MC when we apply the single-particle responses
from data and MC to the individual particles within MC jets that are spatially
matched to reconstructed jets~\citep{spr}:
\begin{equation}
{\cal D}=\frac{\sum{E_i\cdot R_i^{\rm Data}}}{\sum{E_i\cdot R_i^{\rm MC}}},
\end{equation}
where the sums run over each particle $i$ in the MC particle jet, $E_i$ is the
true energy of particle $i$, and  $R_i^{\rm Data}$ and $R_i^{\rm MC}$ are the
single-particle responses in data and MC, respectively.  We define a correction
factor for a jet of flavor $\beta\, (={\rm light\,quark}, {\rm gluon, or}\,
b\,{\rm quark})$ as the ratio of the discrepancy for jets of flavor $\beta$  to
the flavor-averaged discrepancy for jets in $\gamma+$jet events,  $F_{\rm
corr}^\beta={\cal D}^\beta/\left<{\cal D}^{\gamma+\rm jet}\right>$.
Defining the correction this way preserves the standard MC jet energy scale that
is, strictly speaking, only appropriate for the $\gamma+$jet events from which
it is derived.  At the same time, it brings the relative response difference
between jets of flavor $\beta$ and jets in $\gamma+$jet events in MC into
agreement with that in data. The quantity $F_{\rm corr}^\beta-1$ is shown in
Fig.~\ref{fig:flavcorr} as a function of jet $\pt$ and $\eta$ for light-quark,
gluon, and $b$ jets.  The shaded band at $F_{\rm corr}^\beta-1=0$ in each plot
corresponds to the correction for jets in $\gamma+$jet events.  We apply these
correction factors to the light-quark jets and $b$ jets in a \ttbar MC sample
generated with $\mtgen=172.5$ GeV, extract \mtop and \kjes using our analysis
technique, and compare them with the values extracted from the same set of
events without using this correction.  We find shifts of $\Delta\mtop=1.26$ GeV
and $\Delta\kjes=-0.005$ relative to the uncorrected sample.  Repeating this
study on a \ttbar MC sample appropriate for the previous analysis
\citep{meljprl} yields shifts of $\Delta\mtop=1.28$ GeV and
$\Delta\kjes=-0.005$.

\section{Measurement of the Top-Quark Mass}
\label{sec:results}

\begin{figure}
\begin{centering}
\includegraphics[width=0.49\columnwidth]{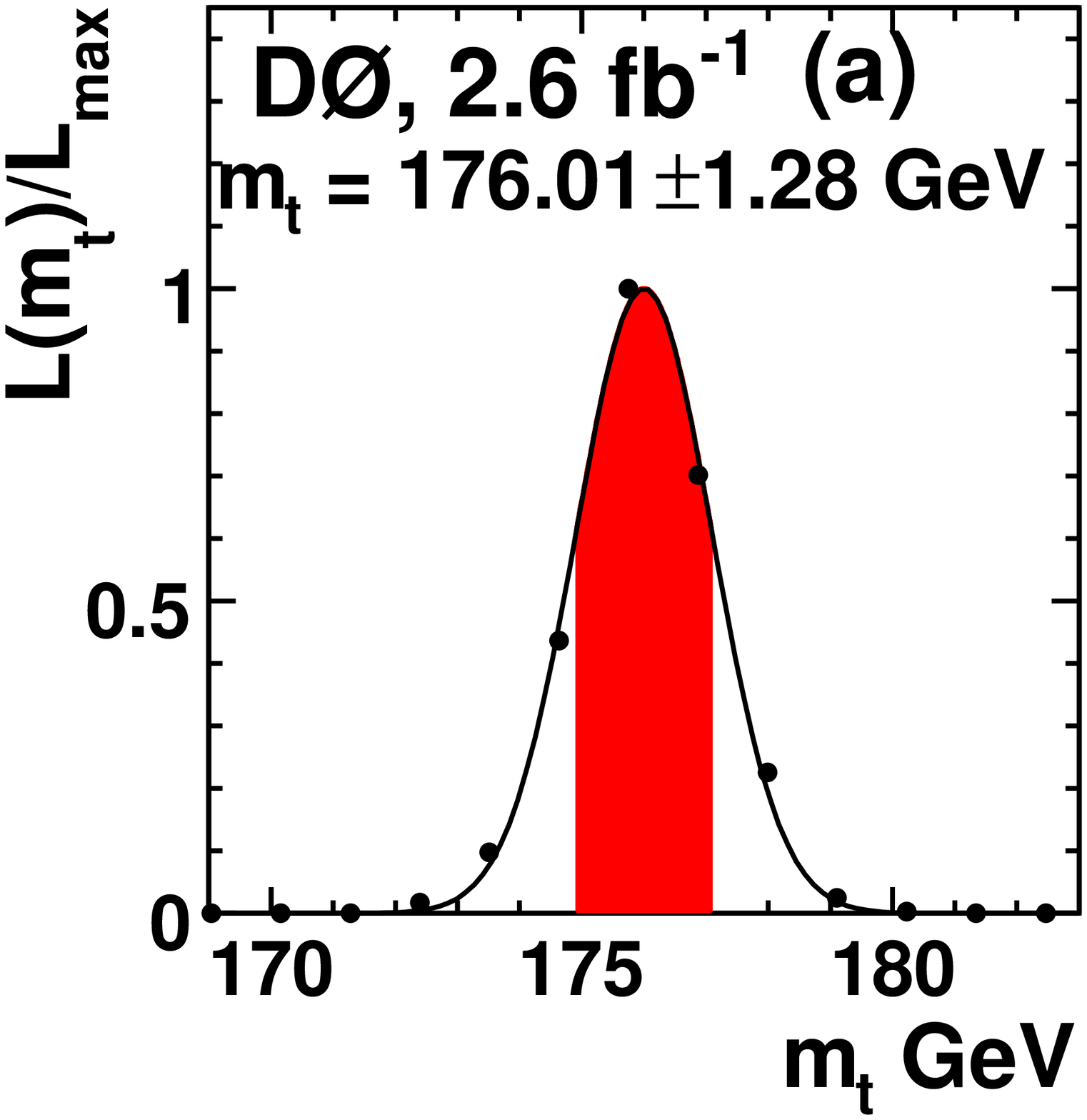}
\includegraphics[width=0.49\columnwidth]{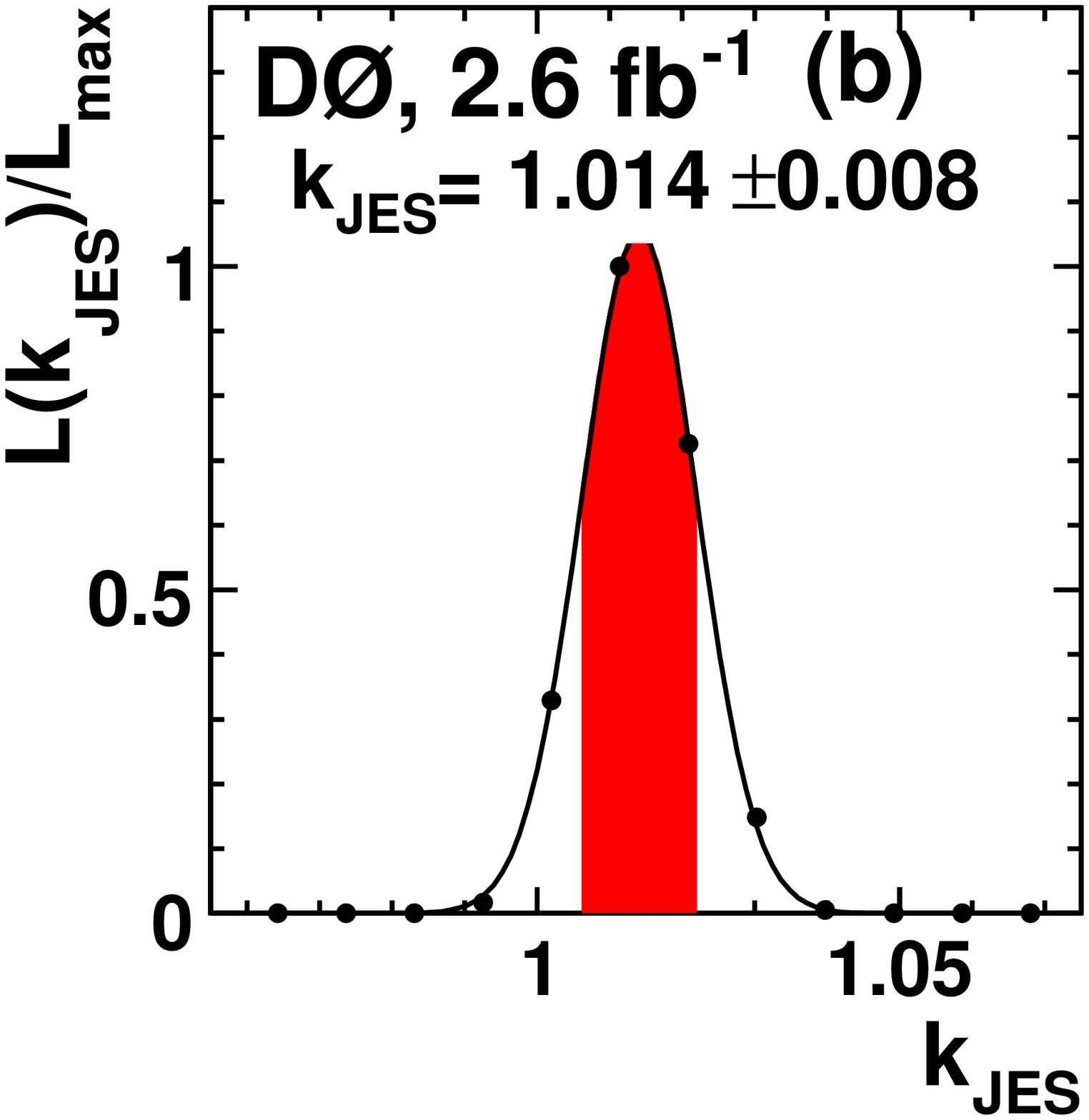}
\par\end{centering}
\caption{\label{fig:like}Calibrated projections of the data likelihoods onto the
(a) \mtop and (b) \kjes axes with 68\% confidence level regions indicated by the
shaded areas.  The values of \mtop and \kjes shown in the figures are after
applying all the corrections described in Sec.~\ref{sec:flavcorr} and
Sec.~\ref{sec:results}. }
\end{figure}

\begin{figure}
\begin{centering}
\includegraphics[width=0.98\columnwidth]{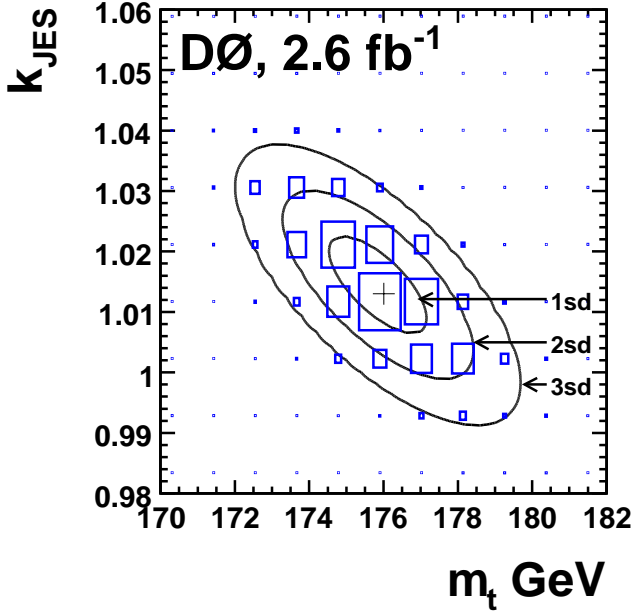}
\par\end{centering}
\caption{\label{fig:like2d}(color online) Fitted contours of equal probability for the
two-dimensional likelihood $L\left(\xt;\, \mtop,\kjes\right)$.}
\end{figure}

\begin{figure}
\begin{centering}
\includegraphics[width=0.49\columnwidth]{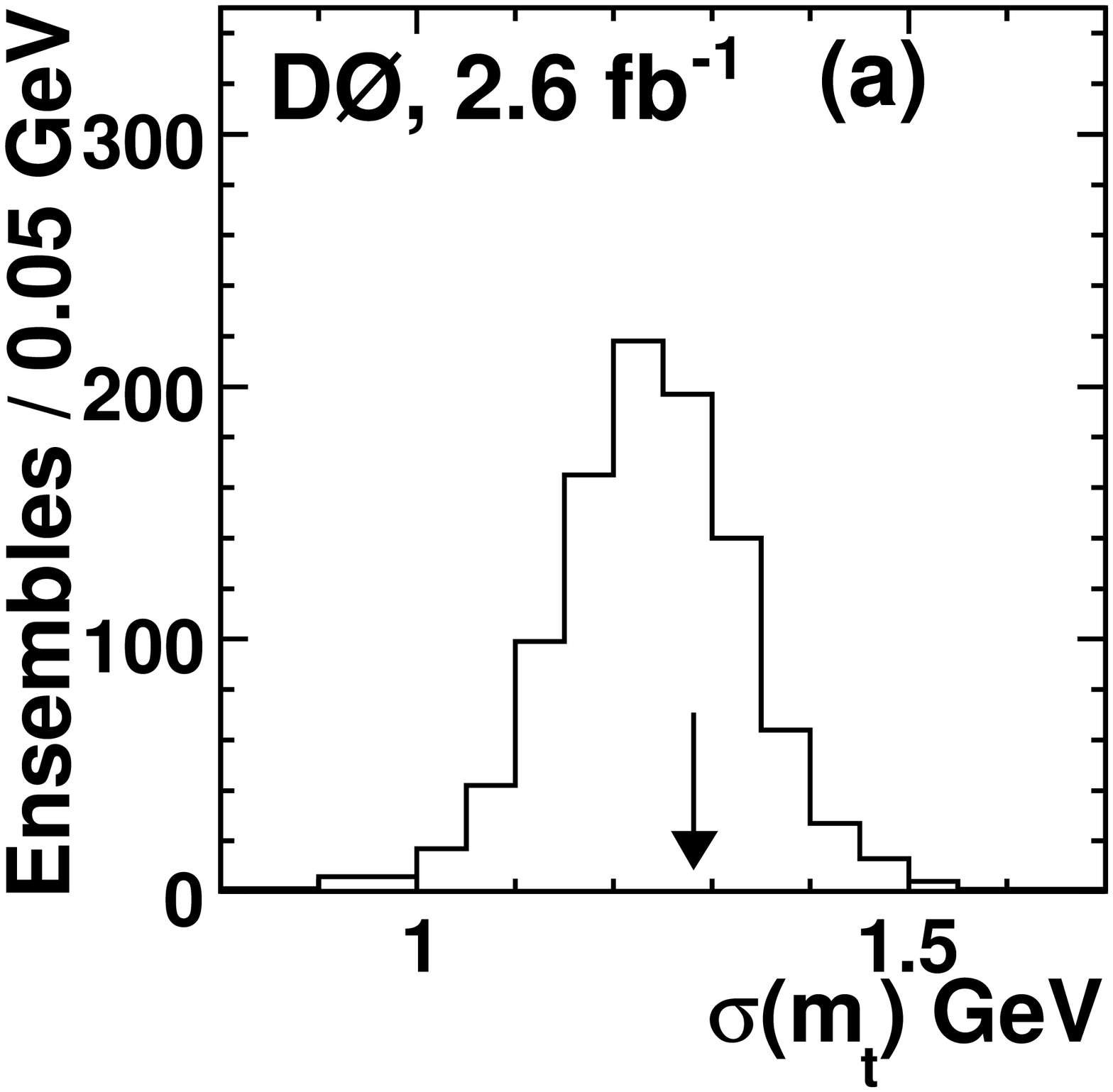}
\includegraphics[width=0.49\columnwidth]{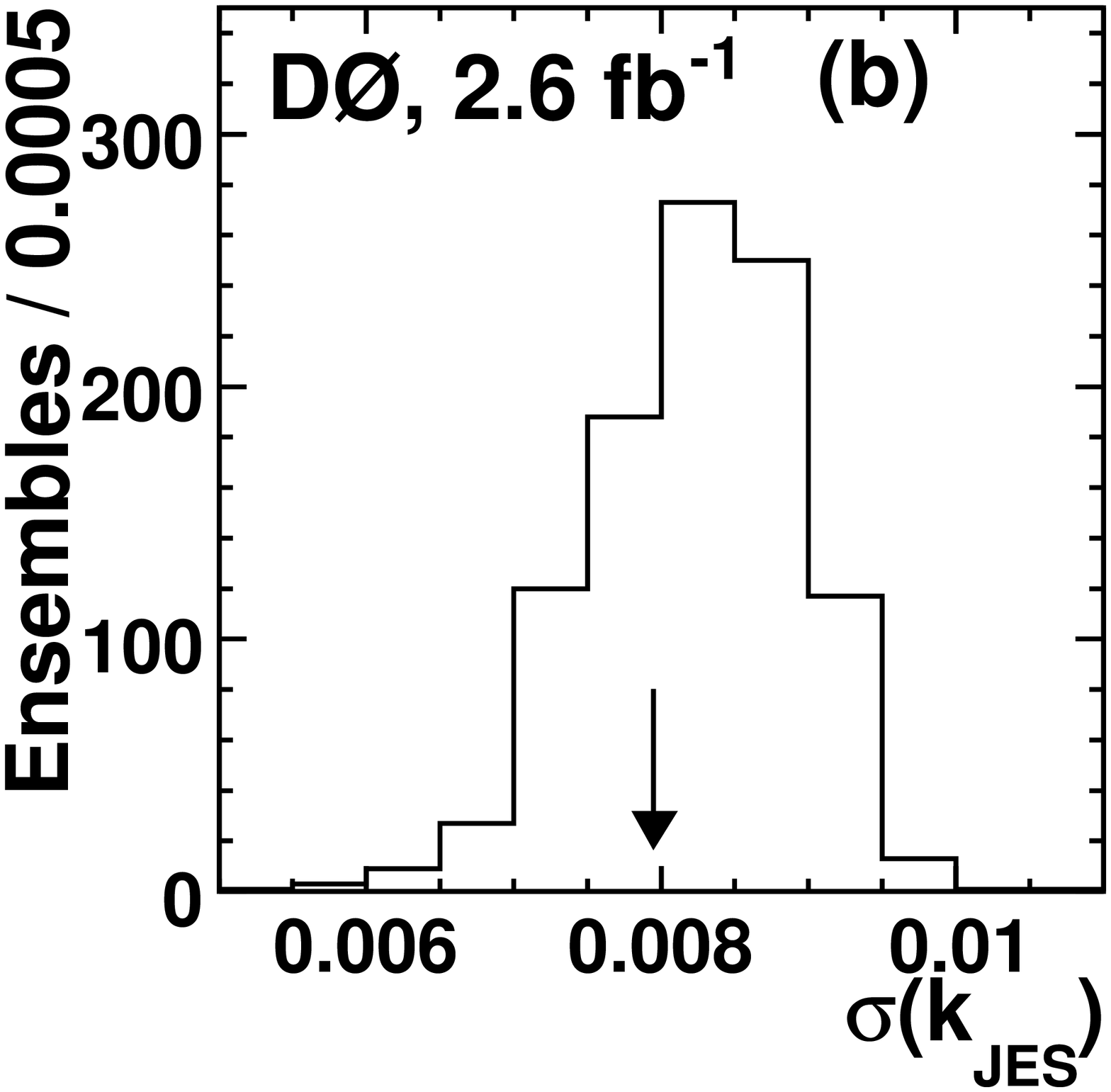}
\par\end{centering}
\caption{\label{fig:err}Expected uncertainty distributions for (a) \mtop and
(b) \kjes determined from 1000 pseudoexperiments performed on the MC \ttbar sample 
for $\mtgen=175$ GeV. The measured uncertainties are indicated by the arrows.}
\end{figure}

The likelihoods $L\left(\xt;\, \mtop\right)$ and $L\left(\xt;\, \kjes \right)$
for the selected data, calculated according to Eq.~(\ref{eq:likeproj_mtop}) and
Eq.~(\ref{eq:likeproj_jes}), respectively, are calibrated by replacing \mtop and
\kjes by parameters fitted to the response plots of Sec. \ref{sec:calib}:
\begin{eqnarray}
\mtop^{\rm calib} & = & \frac{(\mtop-172.5\, {\rm
GeV})-p_0^{\mtop}}{p_1^{\mtop}}+ 172.5\, {\rm GeV},\hspace{10pt}\\
\kjes^{\rm calib} & = & \frac{(\kjes-1)-p_0^{\kjes}}{p_1^{\kjes}}+1,
\end{eqnarray}
where $p_i^{\mtop}$ and $p_i^{\kjes}$ are the parameters of the \mtop and \kjes
response functions shown in Fig.~\ref{fig:calib}(a) and Fig.~\ref{fig:calib}(b),
respectively, and \mtop and \kjes and their uncertainties are extracted from the
mean and RMS values of the calibrated likelihoods shown in
Figs.~\ref{fig:like}(a) and \ref{fig:like}(b).  
The extracted uncertainties for \mtop and \kjes are multiplied by 1.08 and 1.07,
respectively, to correct for deviations of the average pull widths from unity 
(see Sec.~\ref{sec:calib}).
Figure~\ref{fig:like2d} shows the fitted Gaussian
contours of equal probability for the two-dimensional likelihoods as a function
of \mtop and \kjes. We find $\mtop = 174.75\pm1.28({\rm stat+JES})\, {\rm GeV}$
and $\kjes = 1.018\pm0.008(\rm stat)$.  Applying the shifts of
$\Delta\mtop=1.26$ GeV and $\Delta\kjes=-0.005$ described in Sec.
\ref{sec:flavcorr} yields a measured top-quark mass and jet energy scale factor
of
\begin{eqnarray*}
\mtop & = & 176.01\pm1.28({\rm stat+JES})\, {\rm GeV}\\
      & = & 176.01\pm1.01({\rm stat})\pm0.79({\rm JES})\, {\rm GeV},\\
\kjes & = & 1.013\pm0.008(\rm stat).
\end{eqnarray*}
Distributions in expected uncertainties, determined from 1000 pseudoexperiments
performed on the MC \ttbar sample for $\mtgen=175$ GeV, 
are shown in Figs. \ref{fig:err}(a) and \ref{fig:err}(b) for \mtop and \kjes, respectively.
The measured uncertainties, indicated by the arrows, are within the expected range
observed in MC, and do not depend in any appreciable way on the assumed value of $m_t$.

\section{Systematic Uncertainties}
\label{sec:syst}
We evaluate systematic uncertainties for three categories.  The first category,
\emph{modeling of production}, addresses uncertainties in the MC modeling of
\ttbar and \wjets production.  The second category, \emph{modeling of detector},
deals with the uncertainties in jet energy and lepton momentum scales and the
simulation of detector response and any associated efficiencies.  The third
category involves uncertainties in the calibration of both \mtop and the signal
fraction $f$, and a possible bias from the exclusion of multijet events in
MC ensemble studies. The contributions to the systematic uncertainty are
summarized in Table \ref{tab:syst}.  In the first three sections below, we
describe the evaluation of each of these contributions in more detail.  In the
fourth section, we discuss how systematic uncertainties from the previous
analysis of $1$ fb$^{-1}$ of integrated luminosity \citep{meljprl} are updated
to facilitate the combination of the two results presented in Sec.
\ref{sec:combination}.  Except for the \emph{Data-MC jet response difference}
described in Sec. \ref{sec:flavcorrsyst}, all of the systematic uncertainties
described below are calculated prior to the flavor-dependent jet response
corrections of Sec. \ref{sec:flavcorr}.

We adopt the following convention for systematic uncertainties
$\delta\mtop$ in \mtop, and classify them into two types.  The
first type, referred to as the \emph{Type I} uncertainty, is the effect of the
$\pm1$~sd variation of a relevant quantity.  The second type, referred to as the
\emph{Type II} uncertainty, is due to the difference between models.  For Type I
uncertainties, we refer to the central or default value of the measurement as
$\mtop^0$ and to the measurement corresponding to the $1$~sd ($-1$~sd) variation
as $\mtop^+$ ($\mtop^-$).  We compute Type I uncertainties according to
$\delta\mtop=\left|\mtop^+-\mtop^-\right|/2$ if $\mtop^-<\mtop^0<\mtop^+$ or
$\mtop^+<\mtop^0<\mtop^-$ and according to
$\delta\mtop=\max\left(\left|\mtop^+-\mtop^0\right|/2,\left|\mtop^--\mtop^0\right|/2\right)$
if $\mtop^+-\mtop^0$ and $\mtop^--\mtop^0$ have the same sign.  We compute Type
II uncertainties by taking the maximal difference between the models as the $+$
and $-$ systematic variations.

Many of our systematic uncertainties are evaluated by comparing two MC \ttbar
samples generated with the same input mass \mtgen.  For these studies, we use
samples with a value of \mtgen close to the world average of \mtop such as
$172.5$ GeV or $170$ GeV.

\begin{table}
\caption{\label{tab:syst}Summary of systematic uncertainties.}
\vspace{.25cm}
\centering
\begin{tabular}{lc}
\hline
\hline 
Source  & Uncertainty (GeV)\\
\hline
\multicolumn{2}{l}{\textit{Modeling of production:}}\\
\multicolumn{2}{l}{\hspace{12pt}\textit{Modeling of signal:}}\\
\hspace{24pt}Higher-order effects & $\pm0.25$\\
\hspace{24pt}ISR/FSR & $\pm0.26$\\
\hspace{24pt}Hadronization and UE & $\pm0.58$\\
\hspace{24pt}Color reconnection & $\pm0.28$\\
\hspace{24pt}Multiple \ppbar interactions & $\pm0.07$\\
\hspace{12pt}Modeling of background  & $\pm0.16$\\
\hspace{12pt}\wjets heavy-flavor scale factor  & $\pm0.07$\\
\hspace{12pt}Modeling of $b$ jets & $\pm0.09$\\
\hspace{12pt}Choice of PDF & $\pm0.24$\\
\multicolumn{2}{l}{\textit{Modeling of detector:}}\\
\hspace{12pt}Residual jet energy scale  & $\pm0.21$\\
\hspace{12pt}Data-MC jet response difference & $\pm0.28$\\
\hspace{12pt}$b$-tagging efficiency  & $\pm0.08$\\
\hspace{12pt}Trigger efficiency & $\pm0.01$\\
\hspace{12pt}Lepton momentum scale & $\pm0.17$\\
\hspace{12pt}Jet energy resolution & $\pm0.32$\\
\hspace{12pt}Jet ID efficiency & $\pm0.26$\\
\multicolumn{2}{l}{\textit{Method:}}\\
\hspace{12pt}Multijet contamination  & $\pm0.14$\\
\hspace{12pt}Signal fraction  & $\pm0.10$\\
\hspace{12pt}MC calibration  & $\pm0.20$\\
\hline
Total  & $\pm1.02$\\
\hline
\hline
\end{tabular}
\end{table}
\subsection{Modeling of Production\label{sub:physicsmodeling}}

\subsubsection{Higher-Order Effects}
The MC \ttbar samples used to calibrate our measurement are generated using \alpgen
for the hard-scattering process and \pythia for shower evolution and
hadronization (Sec. \ref{sec:mcsamples}).  We compare the LO generator \alpgen
with the next-to-leading order MC generator \mcnlo \citep{mcnlo}, in order to
evaluate possible contributions from higher-order effects such as additional
radiation of hard jets or $gg$ contributions.  We compare \alpgen and \mcnlo MC
\ttbar samples with identical values of \mtgen that both use
\herwig~\citep{herwig} for shower evolution and hadronization.  \herwig is used
in both cases for consistency because \mcnlo can only be used with \herwig
(\alpgen can be used with \pythia or \herwig) and we are not interested in 
comparing different models for shower evolution and hadronization in this study.
Ensemble studies are performed on both samples and the difference in the mean
extracted \mtop from ensembles for the two samples is found to be $\mtop^{\text
\mcnlo}-\mtop^{\text \alpgen}=0.10\pm0.25$ GeV.  Here, as in all the other
systematic sources described below, when a shift in the value of the estimated
parameter is statistically dominated, we replace the shift with its statistical
uncertainty for the estimate of uncertainty.  We, therefore, assign an
uncertainty of $\pm0.25$ GeV as the contribution from this source.

\subsubsection{ISR/FSR}
The uncertainties from this source are in the modeling of additional jets due to
initial and final-state radiation (ISR/FSR).  To evaluate this contribution, we
compare three \pythia samples having identical values of \mtgen, with input
parameters taken from a CDF ISR/FSR study based on the Drell-Yan process
\cite{cdfisr}. The three sets of parameters correspond to a fit to data and
$\pm1$~sd excursions.  Half of the difference between the two excursions
corresponds to a change in \mtop of $0.26\pm0.19$ GeV.

\subsubsection{Hadronization and Underlying Event}
In simulating parton evolution and hadronization, \pythia and \herwig model the
parton showering, hadronization, and underlying event (UE)  differently.  To
estimate the impact of this difference, we compare two MC \ttbar samples with
identical values of \mtgen, using \alpgen for the hard-scattering process, but
one sample using \pythia and the other using \herwig for parton showering and
hadronizaton.  Ensemble studies indicate a difference in the means of the
extracted \mtop to be $\mtop^{\text \pythia}-\mtop^{\text \herwig}=0.58\pm0.25$
GeV. 

\subsubsection{Color Reconnection}
The MC samples used in this analysis do not simulate color reconnection for the
final-state particles~\citep{wickeskands}.  To evaluate the possible effect of
color reconnection on the determination of \mtop, we compare two MC \ttbar
samples with identical values of \mtgen, using \pythia 6.4 tunes {\tt Apro}
and {\tt ACRpro}, which are identical except for the inclusion of  color
reconnection in {\tt ACRpro}.  Ensemble studies of \ttbar events performed on
both samples yield a difference in the means of the extracted \mtop of
$\mtop^{\tt Apro}-\mtop^{\tt ACRpro}=0.26\pm0.28$ GeV.  We take the uncertainty
on this difference and assign $\pm0.28$ GeV as the contribution from this
source.

\subsubsection{Modeling of Jet Mass}
Unlike the jet algorithm used in Run I of Tevatron, the iterative midpoint cone
algorithm used for Run II defines jets of intrinsic mass~\citep{run2jets}. The
effect of inaccuracies in the simulation of jet masses on the top-quark mass
measurement is found to be negligible and is presently ignored.

\subsubsection{Multiple \ppbar Interactions}
Effects from additional \ppbar interactions are simulated by overlaying on MC
events unbiased triggers from random \ppbar crossings. These overlaid events are
then reweighted according to the number of interaction vertices to assure that
the simulation reflects the instantaneous luminosity profile of the data.  To
evaluate the contribution from the uncertainty associated with the reweighting
procedure, we repeat the ensemble studies used to derive the \mtop calibration,
but without the reweighting.  The rederived calibration is applied to
$L\left(\xt;\, \mtop\right)$ for the selected data sample, \mtop is extracted
and compared with the value from the default calibration, and found to shift by
$-0.07$ GeV. This extreme check of the size of this contribution to the
uncertainty shows that our result is not affected significantly by variations in
luminosity.

\subsubsection{Modeling of Background}
\label{sec:bkgmod}
This systematic uncertainty receives contributions from two sources, one based
on the data-MC discrepancy in background-dominated distributions, and a second
from uncertainty in the renormalization scale used to generate the \wjets
samples.  For the first source, we identify distributions in which there is poor
agreement between data and MC in the modeling of background.  Specifically, in
both channels, we examine lepton \pt and the $\eta$ of the jet of lowest \pt in
the  3-jet multiplicity bin. Ensemble studies are performed on a sample of MC
\ttbar events using background events reweighted to match the distributions in
data. The  mean of the extracted \mtop for this sample is found to shift by
$-0.03$ GeV relative to that of the same MC \ttbar events using the default
background events.

The \wjets MC samples used in this analysis (Sec. \ref{sec:mcsamples}) are
generated using identical renormalization and factorization scales of
$\mu=M_W^2+\sum{p_T^2}$ where the sum is over the jets in an event.  To evaluate
the effect of the uncertainty in this scale, we generate two more \wjets MC
samples with modified renormalization and factorization scales of $\mu/2$ and
$2\mu$.  We perform ensemble studies on a \ttbar MC sample using these
modified \wjets samples, and find that the means of the extracted \mtop shift by
$0.13$ GeV ($\mu/2$) and $0.32$ GeV ($2\mu$) relative to the studies using the
default \wjets sample. We take half of the larger excursion and assign $\pm0.16$
GeV as the contribution from this source.

The contributions from the above data-MC discrepancy for the background and from
the uncertainty on the scales are combined in quadrature for a total of
systematic uncertainty of $\pm0.16$ GeV.

\subsubsection{$W$+jets Heavy-Flavor Scale Factor}
\label{sec:hfsf}
The default heavy-flavor content in LO \alpgen MC \wjets (Sec.
\ref{sec:mcsamples}) is increased by a factor of $1.47$ for the
$Wc\overline{c}+$jets and  $Wb\overline{b}+$jets contributions to achieve
agreement with NLO calculations of cross sections that include NLL corrections
based on the \mcfm MC generator \citep{mcfm}.  To evaluate the uncertainty from
this source, we shift this factor up to 1.97 and down to 0.97 and, for each
variation, repeat the ensemble studies described in Sec. \ref{sec:calib} for the
calibration of \mtop, apply this to $L\left(\xt;\, \mtop\right)$ in data, and
re-extract \mtop.  The shifts in \mtop relative to the default value are found
to be $-0.07$ GeV and $0.02$ GeV when the scale factors are shifted up and down,
respectively.  We assign $\pm0.07$ GeV as the contribution from this source to
the uncertainty of \mtop.

\subsubsection{Modeling of $b$ jets}
Possible effects in modeling $b$-quark fragmentation are studied by reweighting
the simulated $t\overline{t}$ events used in the calibration of the measurement
to simulate other choices of $b$-quark fragmentation models for the $b$ jets. 
All the default MC samples used in this analysis consist of events that are
reweighted from the default \pythia $b$-quark fragmentation function (based on
the Bowler model~\cite{bowler}) to a Bowler scheme with parameters tuned to data
collected at the LEP $e^+e^-$ collider~\cite{yvonne}. To evaluate the systematic
uncertainty, these events are reweighted again to account for differences
between LEP and SLAC $e^+e^-$ data~\cite{yvonne}. The ensemble studies of \mtop
are repeated using these reweighted events, the new calibration applied to
$L\left(\xt;\, \mtop\right)$ for data, and \mtop extracted.  \mtop is found to
shift by $0.08$ GeV relative to the default value.

Additional differences in the response of $b$ jets can be expected in the
presence of semileptonic decays of $b$ or $c$-quarks. The incorrect simulation
of semileptonic $b$ and $c$-quark decay branching fractions can therefore lead
to a systematic shift in the extracted value of \mtop. We take an uncertainty of
$\pm0.05$ GeV determined in Ref. \citep{p14prd} as the contribution from this
source.

Combining the two above uncertainties in quadrature gives $\pm0.09$ GeV, which
we assign as the systematic uncertainty for the modeling of $b$ jets.

\subsubsection{Choice of PDF}
We evaluate this systematic uncertainty using a \pythia
MC \ttbar sample that is reweighted to match possible excursions in the PDF
parameters represented by the $20$ CTEQ6M uncertainty PDFs \citep{cteq}.
Ensemble studies are repeated for each of these variants for only \ttbar events,
and the uncertainty evaluated using the following formula \citep{cteq}:
\begin{equation}
\delta\mtop^{\rm PDF}=\frac{1}{2}\left({\displaystyle\sum_{i=1}^{20}[\Delta M(S_{i}^{+})-\Delta M(S_{i}^{-})]^{2}}\right)^{1/2}
\end{equation}
where the sum runs over PDF excursions in the positive ($S_{i}^{+})$ and
negative ($S_{i}^{-}$) directions. $\delta\mtop^{\rm PDF}$ is found to be
$0.24$ GeV.

\subsection{Modeling of Detector}

\subsubsection{Residual $JES$ Uncertainty}
\label{sec:rjes}
The {\sl in situ} jet energy calibration employed in this analysis addresses a
possible global scale difference in JES between data and MC.  Any other
discrepancy, such as a dependence on \pt and $\eta$, can have a systematic
effect on the determination of \mtop. To estimate this, the fractional
uncertainty associated with the standard jet energy correction, derived using
the $\gamma+$jet and dijet samples, is parameterized as a function of $p_{T}$
and $\eta$.  This uncertainty includes statistical and systematic contributions
from both data and MC added in quadrature.  All jet energies in a \ttbar MC
sample are then scaled up by the parameterized uncertainty as a function of
$p_{T}$ and $\eta$.  The parameters are then shifted in such a way that the
average scale shift applied to all jets vanishes.  Ensemble studies are
performed on the default and scaled samples, and the extracted \mtop found to
shift by $0.21$ GeV relative to the default sample.

\subsubsection{Data-MC Jet Response Difference}
\label{sec:flavcorrsyst}
The uncertainties in the flavor-dependent jet response correction for MC events
(described in Sec.~\ref{sec:flavcorr}), used to bring the simulation of
calorimeter response into agreement with that observed in the data, are
associated with uncertainties in single-particle responses in data and MC. To
evaluate the effect of these uncertainties on the value of \mtop, we change the
correction factors by $\pm 1$~sd and apply them to the light jets and $b$ jets
in a \ttbar MC sample.  The value of \mtop is extracted and the mean is found to
shift by $\pm0.28$ GeV relative to the sample corrected using the central
values.

\subsubsection{$b$-Tagging Efficiency}
Discrepancies in the $b$-tagging efficiency between data and MC can lead to a
systematic shift in the extracted \mtop. To evaluate the effect of possible
discrepancies, the tag rate functions for $b$ and $c$ quarks and the mistag rate
function for light quarks are changed by 5\%~\cite{bnim} and 20\%, respectively,
corresponding to the uncertainties on these functions. Ensemble studies for all
\ttbar MC samples are then repeated and the \mtop calibration rederived and
applied to data to extract \mtop.  The result is compared with that from the
default calibration and found to shift by $-0.08$ GeV.

\subsubsection{Trigger Efficiency}
The MC events used in this analysis have associated weights to
simulate the effect of trigger efficiencies.  To evaluate the effect of the
uncertainties in these weights on the top-quark mass, we repeat the
ensemble studies on all \ttbar MC samples with the weights set to unity,
rederive the \mtop calibration, and apply it to the data to extract
\mtop.  The result is found to shift by $-0.01$ GeV.

\subsubsection{Lepton Momentum Scale}
A relative difference in the lepton momentum scale between data and MC can have
a systematic effect on \mtop.  To evaluate this, we first determine the size of
the discrepancy and correct the scale of one \ttbar MC sample.  Ensemble studies
are repeated on the corrected sample and the mean of the extracted \mtop is
found to shift by $0.17$ GeV relative to the default sample.

\subsubsection{Jet Energy Resolution}
Since the jet transfer functions used are derived from MC samples, improper
simulation of jet energy resolution can result in a bias in the extracted
\mtop.  To evaluate a possible bias, ensemble studies are performed using a
\ttbar MC sample with jet energy resolutions degraded by  $1$~sd.  The mean of
the extracted \mtop in this sample is found to shift by $0.32$
GeV~\citep{halftypeI}.

\subsubsection{Jet ID Efficiency}
The uncertainties associated with the scale factors used to achieve data-MC
agreement in jet ID efficiencies are propagated to the measurement of \mtop by
decreasing the jet ID efficiencies in a \ttbar MC sample according to these
uncertainties.  We can only simulate a decrease and not an increase, as
reconstructed jets can be dropped but not created. Ensemble studies indicate
that the mean of the extracted \mtop shifts by $0.26$ GeV relative to that of
the default sample~\citep{halftypeI,systfromp17}.

\subsection{Method}
\subsubsection{Multijet Contamination}
The multijet background is not included in the ensemble studies used to derive the
calibrations described in Sec.~\ref{sec:calib} as we have assumed that
$\Pbkg\gg\Psig$ for such events (see Sec.~\ref{sec:me_method1}), resulting in a
negligible influence on the determination of \mtop. To evaluate possible
systematic effects due to this assumption, we select a multijet-enriched sample
of events from data by inverting the lepton isolation criterion in the event
selections.  We repeat the ensemble studies to derive the \mtop calibration
using the multijet-enriched sample in the sample composition.  The rederived
calibration is applied to data and the extracted \mtop is found to shift by
$0.14$ GeV relative to the default calibration~\citep{systfromp17}.

\subsubsection{Signal Fraction}
The signal fractions determined from data and used in the ensemble studies have
associated statistical uncertainties.  These signal fractions are varied by
their uncertainties, independently for each decay channel, and the ensemble
studies repeated for all MC samples to rederive the \mtop calibration shown in
Fig.~\ref{fig:calib}(a). The new calibrations are then applied to the data and
results compared with those obtained using the default calibration. The
resulting uncertainties in \mtop evaluated by changing the signal fractions in
each decay channel are then added in quadrature and divided by two to obtain a
total of $\pm0.10$ GeV.

\subsubsection{MC Calibration}
We estimate the effect of the statistical uncertainties associated with the
offset and slope parameters determined from the fit to the response plot shown
in Fig. \ref{fig:calib}(a).  To estimate this uncertainty, we change these two
parameters, one at a time, by their uncertainties, and apply the modified
calibration to the data to extract \mtop, and calculate the difference relative
to the \mtop extracted using the default calibration.  We combine, in
quadrature, the differences in \mtop resulting from such changes in each
parameter, and find an uncertainty of $\pm0.20$ GeV.

\subsection{Treatment of Systematic Uncertainties in Previous Analysis}
\label{sec:p17syst}
To facilitate the combination of the new measurement with the previous one, we
have updated the systematic uncertainties presented in Table I of
Ref.~\citep{meljprl}.  All of the uncertainties in this table are unchanged,
except for the uncertainties in the modeling of signal and the relative
$b$/light-quark response ratio.  The uncertainty for the modeling of signal in
the previous analysis is replaced with one from the current analysis, which
includes contributions from uncertainties in the modeling of higher-order
effects, ISR/FSR, hadronization and underlying event, color reconnection, and
multiple hadron interactions. The uncertainty on $b$/light-quark response is
replaced with that associated with differences in jet response in data and MC
for the current analysis (see also Sec.~\ref{sec:flavcorr}). The uncertainty in
the modeling of background in Table I of Ref.~\citep{meljprl} is the sum in
quadrature of (i) the uncertainty in the heavy-flavor scale factor, and (ii) the
uncertainty associated with discrepancies between data and MC background
distributions.  Since the uncertainty on the renormalization and factorization
scale was not evaluated in the previous analysis, we include the additional
contribution described in the second part of Sec. \ref{sec:bkgmod}.  We also
evaluate the uncertainty associated with the flavor-dependent jet-response
correction factors appropriate for the previous analysis, using the procedure
described in Sec. \ref{sec:flavcorrsyst}.  We find the mean of the extracted
\mtop shifts by $0.13$ GeV ($-0.22$ GeV) relative to the sample corrected with
the central values when we change the correction factors by $1$~sd ($-1$~sd). 
We assign $\pm0.22$ GeV as the contribution from this source.  Adding the
contributions from all sources in quadrature gives a total of $\pm0.97$ GeV.

\section{Result of the Current Measurement}
\label{sec:currentresult}
We measure the mass of the top quark in \ttbar lepton$+$jets events using a
matrix element method that combines an {\sl in situ} jet energy calibration with
additional information from the standard jet energy scale derived from
$\gamma+$jet and dijets samples.  Using data corresponding to 2.6 fb$^{-1}$ of
integrated luminosity  collected by the D0 experiment from Run II of the
Tevatron collider, we extract the value: 
\[
\mtop = 176.01\pm1.01({\rm stat})\pm0.79({\rm JES})\pm 1.02({\rm syst})\, {\rm GeV},
\]
or $\mtop = 176.01\pm1.64$ GeV.
\par\vspace{5mm}
\section{Combination with the Previous Measurement}
\label{sec:combination}
Our result from a previous measurement using the same analysis technique, and
based on earlier data corresponding to 1 fb$^{-1}$ of integrated luminosity, is
$\mtop = 171.5\pm1.76({\rm stat+JES})\pm1.1({\rm syst})\, {\rm GeV}$
\citep{meljprl}.  Applying the shift of $\Delta\mtop=1.28$ GeV described in Sec.
\ref{sec:flavcorr}, and using updated systematic  uncertainties described in
Sec.~\ref{sec:p17syst}, yields
\[
\mtop = 172.74\pm1.44({\rm stat})\pm1.05({\rm JES})\pm0.97({\rm syst})\, {\rm GeV},
\]
or $\mtop=172.74\pm 2.03$ GeV.

We combine the two measurements using the BLUE method \citep{blue1,blue2} to get
a result equivalent to 3.6 fb $^{-1}$ of integrated luminosity.  The combined
value of the mass is
\[
\mtop = 174.94\pm0.83({\rm stat})\pm 0.78({\rm JES})\pm 0.96({\rm syst})\, {\rm GeV},
\]
or $\mtop=174.94\pm 1.49$ GeV.  The procedure we follow uses the same
method and classes of uncertainty as used by the Tevatron Electroweak Working
Group \citep{tevewwg} in combining individual measurements for Tevatron
averages of the top-quark mass.
 
\section*{Acknowledgemements}
%
We thank the staffs at Fermilab and collaborating institutions,
and acknowledge support from the
DOE and NSF (USA);
CEA and CNRS/IN2P3 (France);
FASI, Rosatom and RFBR (Russia);
CNPq, FAPERJ, FAPESP and FUNDUNESP (Brazil);
DAE and DST (India);
Colciencias (Colombia);
CONACyT (Mexico);
KRF and KOSEF (Korea);
CONICET and UBACyT (Argentina);
FOM (The Netherlands);
STFC and the Royal Society (United Kingdom);
MSMT and GACR (Czech Republic);
CRC Program and NSERC (Canada);
BMBF and DFG (Germany);
SFI (Ireland);
The Swedish Research Council (Sweden);
and
CAS and CNSF (China).
%


\end{document}